\documentclass{article}
\usepackage{amssymb}

\usepackage{amsmath}
\usepackage{graphics}
\usepackage{epsf}
\usepackage{rotating}

\input{tcilatex}

\begin{document}

\title{Discrete integrable systems and deformations of associative algebras}
\author{B.G.Konopelchenko \\
\\
Dipartimento di Fisica, Universita del Salento \\
and INFN, Sezione di Lecce, 73100 Lecce, Italy }
\maketitle

\begin{abstract}
Interrelations between discrete deformations of the structure
constants for associative algebras and discrete integrable systems
are reviewed. A theory of deformations for associative algebras is
presented. Closed left ideal generated by the elements
representing the multiplication table plays a central role in this
theory. Deformations of the structure constants are generated by
the Deformation Driving Algebra and governed by the central system
of equations. It is demonstrated that many discrete equations like
discrete Boussinesq equation, discrete WDVV equation, discrete
Schwarzian KP and BKP equations, discrete Hirota-Miwa equations
for KP and BKP hierarchies are particular realizations of the
central system. An interaction between the theories of discrete
integrable systems and discrete deformations of associative
algebras is reciprocal and fruitful. An interpretation of the
Menelaus relation (discrete Schwarzian KP equation), discrete
Hirota-Miwa equation for KP hierarchy, consistency around the cube
as the associativity conditions and the concept of gauge
equivalence, for instance, between the Menelaus and KP
configurations are particular examples.
\end{abstract}

\bigskip


 \bigskip

\section{Introduction}

\ Theory of solitons and modern theory of deformations of associative
algebras have almost the same age, a little above forty. Geographically they
were born pretty near to each other: one in the Princeton university [1,2]
and another in the Pennsylvania university [3,4]. Moreover both these
theories have used one common principal concept, namely, the concept of
deformation. An idea of isospectral deformations was one of the first basic
ideas in the theory of integrable equations [5]. Then, the dressing method
[6], bi-Hamiltonian structures [7] as well as the Backlund and Darboux
transformations represent particular realizations of the various classes of
deformations for solutions of solitons equations (see e.g.[8,9] and
references therein). On the other hand, one of the approaches to the
deformation theory of associative algebras, proposed in [4] was ''... to
take the point of view that the objects being deformed are not merely
algebras, but essentially algebra with a fixed basis'' and to treat '' the
algebraic set of all structure constants as parameter space for deformation
theory''.

\ In spite of this sharing of the idea of deformation the theory of
integrable equations and deformation theory for associative algebras for the
first twenty five years have been developed independently, without any
interconnection and influence of one to another. An apparent difference
between the basic objects in these theories, i.e. between the dependent \
variables in the nonlinear equations and structure constants of associative
algebras, seemed to be so big that, it was thought, these theories cannot
have anything in common.

\ The situation has changed drastically in the beginning of nineties with
the discovery of Witten [10] and Dijkgraaf-Verlinde-Verlinde [11]. They
demonstrated that the function F which defines the correlation function $%
\langle \Phi _{j}\Phi _{k}\Phi _{l}\rangle =\frac{\partial ^{3}F}{\partial
x^{j}\partial x^{k}\partial x^{l}}$ etc in the perturbed two-dimensional
topological field theory obeys the system of equations

\begin{equation}
\sum_{s,t=1}^{N}\frac{\partial ^{3}F}{\partial x^{j}\partial x^{k}\partial
x^{s}}g^{st}\frac{\partial ^{3}F}{\partial x^{t}\partial x^{m}\partial x^{l}}%
=\sum_{s,t=1}^{N}\frac{\partial ^{3}F}{\partial x^{l}\partial x^{k}\partial
x^{s}}g^{st}\frac{\partial ^{3}F}{\partial x^{t}\partial x^{m}\partial xj}
\end{equation}
where j,k,l,m=1,2,...,N and $g^{st}$ are constants. In terms of $C_{jk}^{l}$
defined as

\begin{equation}
C_{jk}^{l}=\sum_{m=1}^{N}g^{lm}\frac{\partial ^{3}F}{\partial x^{j}\partial
x^{k}\partial x^{m}}
\end{equation}
the WDVV equation (1) is of the form

\begin{equation}
\sum_{m=1}^{N}C_{jk}^{m}(x)C_{ml}^{n}(x)=%
\sum_{m=1}^{N}C_{kl}^{m}(x)C_{jm}^{n}(x)
\end{equation}
which is nothing but the condition of associativity for the structure
constants $C_{jk}^{l}$ of the N-dimensional \ algebra of primary fields $%
\Phi _{j}$ [10,11]. Thus, each solution $F(x)$ of the WDVV equation
describes a deformation of the structure constants by the formula (2).

This result has provided us with the remarkable realization of the
Gerstenhaber's approach mentioned above. On the other hand it has revealed a
striking connection between the theory of deformations of associative
algebras and nonlinear partial differential equations (PDEs).

\ WDVV equation (1) and the formulae (2,3) have been immediately interpreted
and formalized by Dubrovin [12,13] as the theory of Frobenius manifolds. It
provides us with the classes of deformations of the so-called Frobenius
algebras. An extension of this approach to general algebras and
correspoinding F-manifolds has been given in [14]. It was shown within these
theories that not only WDVV equation, but many other integrable systems of
nonlinear PDEs both dispersionless and dispersive describe deformations of
associative algebras (see e.g. [13, 15-18]). In few years the
interconnection between the theories of Frobenius and F-manifolds on one
side and the theory of integrable nonlinear PDEs on the other has been well
established.

\ An alternative approach to deformations of structure constants for
associative algebras proposed recently in the papers [19-24] has allowed us
to construct wider classes of deformations. They include the coisotropic
[19,20], quantum [21], discrete [22,23] deformations as well as general
deformations generated by the so-called Deformation Driving Algebra (DDA)
[24]. These classes of deformations are governed by dispersionless,
dispersive, discrete and difference integrable systems with some well-known
integrable equations among them. One of the characteristic features of the
method developed in [19-24] is that it allows us to construct different
classes of deformations, for instance, coisotropic, quantum and discrete
deformations of the same algebra just choosing different DDAs.

\ Theory of integrable systems is nowadays a well-developed and rich theory
which includes a vast variety of nonlinear ordinary and partial differential
equations and discrete equations (see e.g.[25-28]). Theory of discrete
integrable systems form a very important branch of the whole theory. Toda
lattice [29] was the first such equation studied by the inverse scattering
method [30,31]. After that several methods to construct and solve integrable
differential-difference and discrete equations have been developed ( see
e.g. [32-40]). Pretty soon it became clear that some of the discrete
integrable equations play a fundamental role in the whole theory: they are
the generating equations for the infinite hierarchies of continuous
integrable equations and encode the basic algebraic structures associated
with these hierarchies ( see e.g. [35-41]). Discrete integrable equations
have served also as the basic tool in the formulation and study of discrete
geometry (see [42-44]) and discrete complex analysis [45-47]. Moreover some
of discrete equations are directly connected with basic theorems, like the
Menelaus theorem, of the classical geometry [48-51]. Due to their
significance discrete integrable systems certainly merit a profound study
from all possible viewpoints.

\ Our principal goal here is to discuss recent results on the interrelations
between some basic discrete integrable systems and discrete deformations of
structure constants of associative algebras. We will present first a general
theory of deformations of associative algebras generated by the Lie algebra
DDA and governed by the so-called central system \ (CS) of equations. \ We
will show how discrete Boussinesq and WDVV equations, discrete Schwarzian
Kadomtsev-Petviashvili (DSKP) equation, discrete bilinear Hirota-Miwa
equations for KP and BKP hierarchies, discrete Darboux system and other
discrete systems arise as the particular versions of the CSs which govern
discrete and difference deformations of associative algebras. \ Difference
CS has a simple geometrical meaning of vanishing '' discrete'' Riemann
curvature tensor with the structure constants $C_{jk}^{l}$ playing the role
of ''discrete'' Christoffel symbols.

\ Such an interpretation of discrete equations allows us to understand
better the algebraic backgrounds of the theory of discrete integrable
systems and associated constructions in discrete geometry. \ We will show,
for instance, \ that the DSKP equation or Menelaus relation and Hirota-Miwa
bilinear equation for KP hierarchy are just the associativity conditions for
the structure constants of certain algebras. The consistency around cube and
multi-dimensional consistency discussed in discrete geometry also has a
meaning of conditions of associativity for elements of algebras. On the
other hand the transfer of the old concept of gauge equivalency from the
theory of integrable systems (see e.g. [25-28]) to deformation theory
introduces the notion of gauge equivalence classes of deformations. In
geometrical terms this leads to the notion of the gauge equivalency between
geometrical configurations. \ Menelaus and KP six points configurations on
the plane represent an important example of such situation.

\ Interpretation of discrete integrable systems as equations governing
deformations of associative algebras provides us also with a method for
construction of integrable discretizations of integrable PDEs. This problem
has been intensively discussed for many years and several methods have been
proposed ( see e.g. [32-40]). In our approach an integrable discretization
is just the change of DDA from the Heisenberg algebra to the algebra of
shifts for an associative algebra in the given basis, i.e. for the same
structure constants. Interinfluence of the theories of integrable systems
and deformations of associative algebras is revealed to be rather fruitful.

\ The paper is organized as follows. In section 2 we briefly review some
well-known integrable equations, namely, the discrete Korteweg-de Vries
(KdV) equation, discrete Schwarzian KdV (DSKdV) equation, DSKP equation,
discrete Hirota-Miwa equation and \ their connection with the Backlund and
Darboux transformations for continuous integrable equations. We discuss also
the indications on the possible role of associative algebras in these
constructions. In section 3 we present a general theory of deformations of
structure constants for associative algebras. A closed left ideal generated
by elements representing the multiplication table plays the central role in
this consruction. Deformations of structure constants are generated by the
DDA and are governed by the corresponding CS. \ A subclass of deformations,
the so-called integrable deformations , is discussed in section 4. The CS
for such deformations has a geometrical meaning of vanishing discretized
Riemann curvature tensor. Discrete deformations of the three-dimensional
associative algebra and discrete Boussinesq equation are studied in section
5. Next section 6 is devoted to the discrete and semi-discrete versions of
the WDVV equation. Deformations generated by the three-dimensional Lie
algebras and corresponding discrete mappings are considered in section 7.
Discrete versions of the oriented associativity equation are discussed in
section 8. Discrete deformations of algebras for which the product of only
distinct elements of the basis is defined are studied in section 9.
Deformations of the three-dimensional algebras of such type, Menelaus
configurations and deformations are considered in section 10. In section 11
the KP configurations, discrete KP deformations and their gauge equivalence
to the Menelaus configurations and deformations are analyzed. Section 12 is
devoted to the multidimensional extensions of the Menelaus and KP
configurations and deformations. Deformations governed by the discrete
Darboux system and discrete BKP Hirota-Miwa equation are considered in
section 13.

\section{Backlund-Darboux transformations, discrete integrable systems and
algebras behind them.}

\ Associative algebras show up in various branches of the theory of
integrable continuous and discrete systems. One of the simplest and,
probably, algebraically the most transparent way to establish a connection
between the continuous and discrete integrable equations and to reveal a
possible role of associative algebras in their constructions is provided by
Backlund and Darboux transformations.

Backlund transformations (BTs) and Darboux transformations (DTs) are the
discrete transformations ( depending on parameters) which act on the variety
of solutions of given integrable PDE (see e.g. [52,53]). They commute and,
as the consequence, one has the algebraic relations between several
solutions of the original PDE which are usually refered as the nonlinear
superposition formulae (NSF). Due to the commutativity one can treat BTs and
DTs as the shifts on the lattice and the corresponding NSF take the form of
the discrete equation on this lattice (see e.g. [54,34]).

Probably, the first demonstration of the efficiency of this scheme is
associated with the sine-Gordon equation

\begin{equation}
\varphi _{xy}=\sin \varphi
\end{equation}
where $\varphi _{x}=\frac{\partial \varphi }{\partial x}$ etc. Introduced
and well studied within the classical differential geometry of surfaces in $%
R^{3}$ (see e.g. [52,53]) more than century ago, this equation has been
recognized as integrable by the inverse scattering transform (IST) method in
1973 [55,56]. BT $\varphi \rightarrow \varphi _{1}=B_{a_{1}}\varphi $ for
the sine-Gordon equation is defined by the relations [57,52,53]

\begin{equation}
\frac{1}{2}\left( \varphi _{1}-\varphi \right) _{x}=a_{1}\sin \left( \frac{%
\varphi _{1}+\varphi }{2}\right) ,\quad \frac{1}{2}\left( \varphi
_{1}+\varphi \right) _{y}=\frac{1}{a_{1}}\sin \left( \frac{\varphi
_{1}-\varphi }{2}\right)
\end{equation}
where $a_{1}$ is an arbitrary parameter. BTs (5) with different a commute $%
B_{a_{1}}B_{a_{2}}=B_{a_{2}}B_{a_{1}}$. This leads to the following NSF

\begin{equation}
\tan \left( \frac{\varphi _{12}-\varphi }{4}\right) =\frac{a_{2}+a_{1}}{%
a_{2}-a_{1}}\tan \left( \frac{\varphi _{2}-\varphi _{1}}{4}\right)
\end{equation}
where $\varphi _{12}=B_{a_{1}}B_{a_{2}}\varphi $. This pure algebraic
relation between four solutions of \ equation (4) is a very useful one. \ It
has allowed to calculate all multisoliton solutions of the sine-Gordon
equation many years before the discovery of the IST method [58]. Then, due
to the commutativity of BTs one can treat $B_{a_{1}}$and $B_{a_{2}}$ as the
shifts $T_{1}$ and $T_{2}$ of the solution $\varphi $ at fixed x and y: $%
T_{1}\varphi (x,y)=\varphi _{1}(x,y),T_{2}\varphi (x,y)=\varphi
_{2}(x,y),T_{1}T_{2}\varphi (x,y)=\varphi _{12}(x,y).$ Enumerating \ the
family of solutions of equation (4) obtained by all compositions of BTs by
two integers $n_{1}$, $n_{2}$ such that $T_{1}\varphi (n_{1},n_{2})=\varphi
(n_{1}+1,n_{2}),T_{2}\varphi (n_{1},n_{2})=\varphi (n_{1},n_{2}+1)$etc , one
rewrites the NSF (6) in the form of the discrete equation

\begin{equation}
\tan \left( \frac{T_{1}T_{2}\varphi -\varphi }{4}\right) =\frac{a_{2}+a_{1}}{%
a_{2}-a_{1}}\tan \left( \frac{T_{2}\varphi -T_{1}\varphi }{4}\right) .
\notag \\
\end{equation}

This equation represents the discretization of equation (4). Remarkably, it
coincides (up to some trivial redefinitions) with the integrable
discretization of the sine-Gordon equation proposed by Hirota in [33] within
a completely different method.

\ There are many examples of such type. For the celebrated KdV equation

\begin{equation}
\ u_{t}+u_{xxx}-6u_{x}u=0  \notag \\
\end{equation}
the spatial part of BT $u\rightarrow u_{1}$ is given by [59]

\begin{equation}
\left( u_{1}+u\right) _{x}+(u_{1}-u)\sqrt{\alpha _{1}^{2}-2(u_{1}+u)}=0
\end{equation}
where $\alpha _{1}$ is a parameter. The corresponding NSF of term of the
potential V defined by $u_{x}=V$ is [59]

\begin{equation}
\ V_{12}-V=\frac{(\alpha _{1}+\alpha _{2})(V_{2}-V_{1})}{\alpha _{1}-\alpha
_{2}+V_{2}-V_{1}}.
\end{equation}

Interpreted as the discrete equation this NSF is exactly the discrete KdV
equation

\begin{equation}
\left( \alpha _{1}-\alpha _{2}+(T_{2}-T_{1})V\right) \left( \alpha
_{1}+\alpha _{2}-(T_{1}T_{2}-1)V\right) =\alpha _{1}^{2}-\alpha _{2}^{2}
\notag \\
\end{equation}
introduced in [36,37] within the direct linearization aprroach. This
discrete KdV equation is the generating equation for the whole KdV hierarchy.

In the same manner one can construct the discrete modified KdV equation and
discrete Schwarzian KdV equation [36-38]. The latter one is [38]

\begin{equation}
\frac{(\Phi -T_{1}\Phi )(T_{1}T_{2}-T_{2}\Phi )}{(T_{1}\Phi -T_{1}T_{2}\Phi
)(T_{2}\Phi -\Phi )}=\frac{q^{2}}{p^{2}}
\end{equation}
where p and q are arbitrary real parameters.

Similar results are valid also for 1+1-dimensional integrable systems of
PDEs. For instance, for the AKNS hierarchy [60] the first member of which is
the system

\begin{equation}
iq_{t}+q_{xx}+2q^{2}r=0,\quad ir_{t}-r_{xx}-2r^{2}q=0
\end{equation}
one has two elementary BTs [61] (spatial parts)

\begin{equation}
B_{\alpha }^{1}:iq^{\prime}_{x}-\frac{1}{2}q^{\prime2}r+2\alpha q^{\prime}
+2q=0,\quad ir_{x}+\frac{1}{2}r^{2}q^{\prime} -2\alpha r-2r^{\prime} =0,
\end{equation}

\begin{equation}
B_{\beta }^{2}:iq_{x}-\frac{1}{2}q^{2}r^{\prime} +2\beta q+2q^{\prime}
=0,\quad ir^{\prime} _{x}+\frac{1}{2}r^{\prime 2}q-2\beta r^{\prime} -2r=0
\end{equation}
where $\alpha $ and $\beta $ are arbitrary parameters. NSF for these BTs
consists of the following two equations [61]

\begin{equation}
q_{12}=q+\frac{2(\alpha -\beta )}{\frac{r_{2}}{2}+\frac{2}{q_{1}}},\quad
r_{12}=r-\frac{2(\alpha -\beta )}{\frac{q_{1}}{2}+\frac{2}{r_{2}}}.
\end{equation}

Again it represents the discrete integrable AKNS system (10) and generates
the whole AKNS hierarchy. Note that this NSF implies the discrete equation ($%
u=\frac{q}{2},v=\frac{r}{2})$

\begin{equation}
(T_{1}T_{2}u-u)(T_{2}v+\frac{1}{T_{1}u})+(T_{1}T_{2}v-v)(T_{1}u+\frac{1}{%
T_{2}v})=0.
\end{equation}

Algebraic NSFs for 1+1-dimensional integrable equations typically contain
four solutions and consequently their discrete integrable versions usually
are the four points relations on a lattice.

\ For 2+1-dimensional integrable PDEs the situation is quite different. For
instance, for the KP equation the analog of the NSF (8) contains derivatives
and more BT connected solutions are required in order to get a pure
algebraic NSF. We will consider here the KP equation and hierarchy as
illustrative example. We will also use the technique based on DT to derive
NSF (see [48]).

We start with the standard linear problem

\begin{equation}
\psi _{y}=\psi _{xx}+u\psi
\end{equation}
and adjoint linear problem

\begin{equation}
-\psi _{y}^{\ast }=\psi _{xx}^{\ast }+u\psi ^{\ast }
\end{equation}
for the KP hierarchy. Standard DTs are given by ( see e.g. [8])

\begin{equation}
D_{i}:T_{i}\psi =\psi _{x}-\frac{\psi _{ix}}{\psi _{i}}\psi ,\quad
T_{i}u=u+2(\ln \psi _{i})_{xx,} \quad i=1,2,3
\end{equation}
where $\psi _{i}$,$i=1,2,3$ are independent solutions of the problem (15)
with the original potential u. Subsequent action of two DTs (17) with
distinct $\psi _{i}$ is of the form

\begin{equation}
T_{k}T_{i}\psi =T_{k}\psi _{x}-\frac{T_{k}\psi _{ix}}{T_{k}\psi _{i}}%
T_{k}\psi ,\quad i\neq k
\end{equation}
where

\begin{equation}
T_{k}\psi _{i}=\psi _{ix}-\frac{\psi _{kx}}{\psi _{k}}\psi _{i},\quad i\neq
k.
\end{equation}

DTs commute: $T_{i}T_{k}=T_{k}T_{i}.$ As the consequence, the elimination of
$\psi _{i},\psi _{ix}$ and $T_{k}\psi _{i}$ from the formulae (17), (18)
gives the algebraic NSF [62]

\begin{equation}
T_{1}\left( \frac{(T_{2}-T_{3})\psi }{\psi }\right) +T_{2}\left( \frac{%
(T_{3}-T_{1})\psi }{\psi }\right) +T_{3}\left( \frac{(T_{1}-T_{2})\psi }{%
\psi }\right) =0.
\end{equation}

In a similar manner one obtains the NSF for the wavefunction $\psi ^{\ast }$
of the adjoint problem (16):

\begin{equation}
\frac{(T_{1}-T_{2})\psi ^{\ast }}{T_{1}T_{2}\psi ^{\ast }}+\frac{%
(T_{2}-T_{3})\psi ^{\ast }}{T_{2}T_{3}\psi ^{\ast }}+\frac{(T_{3}-T_{1})\psi
^{\ast }}{T_{3}T_{1}\psi ^{\ast }}=0.
\end{equation}

For binary DT [8] $\psi $ transforms as given in (17) and

\begin{equation}
T_{i}\psi ^{\ast }=-\frac{\Phi _{i}}{\psi _{i}},\quad i=1,2,3
\end{equation}
where $\Phi _{i}\doteqdot \Phi (\psi _{i},\psi ^{\ast })$ and the bilinear
potential $\Phi \doteqdot \Phi (\psi ,\psi ^{\ast })$ is defined by

\begin{equation}
\Phi _{x}=\psi \psi ^{\ast },\quad \Phi _{y}=\psi ^{\ast }\psi _{x}-\psi
\psi _{x}^{\ast }.
\end{equation}

For the potential $\Phi $ the binary DTs are of the form

\begin{equation}
T_{i}\Phi =\Phi -\Phi _{i}\frac{\psi }{\psi _{i}},\quad T_{k}T_{i}\Phi
=T_{k}\Phi -T_{k}\Phi _{i}\cdot \frac{T_{k}\psi }{T_{k}\psi _{i}},\quad
i,k=1,2,3;i\neq k
\end{equation}
where

\begin{equation}
T_{k}\Phi _{i}=\Phi _{i}-\Phi _{k}\frac{\psi _{i}}{\psi _{k}}.
\end{equation}

Since $T_{i}T_{k}\Phi =$ $T_{k}T_{i}\Phi $ the elimination of $\Phi
_{i},\psi _{i}$ and $\psi _{ix}$ from the above formulae gives rise to the
following NSF

\begin{equation}
\frac{(T_{1}\Phi -T_{1}T_{2}\Phi )(T_{2}\Phi -T_{2}T_{3}\Phi )(T_{3}\Phi
-T_{3}T_{1}\Phi )}{(T_{1}T_{2}\Phi -T_{2}\Phi )(T_{2}T_{3}\Phi -T_{3}\Phi
)(T_{3}T_{1}\Phi -T_{1}\Phi )}=-1.
\end{equation}

\ Due to the commutativity of DTs one can interpret their action as the
shifts on the lattice $T_{1}\Phi (n_{1},n_{2},n_{3})=\Phi
(n_{1}+1,n_{2},n_{3})$ etc and, consequently, the NSFs (20), (21) and (26)
represent the discrete equations on the lattice. All of them contain six
points of the lattice in contrast to the 1+1-dimensional case. Under the
constraint $\Phi _{23}=\Phi $ equation (26) is reduced to the discrete
Schwarzian KdV equation (9) [48].

\ Discrete equations (20), (21), (26) are fundamental equations for the KP
hierarchy. They generate the whole hierarchies. For instance, equation (26)
is the generating equation for the Schwarzian KP hierarchy. Moreover it has
a beautiful geometrical meaning in connection with the classical Menelaus
theorem [48].

\ Discrete equations (20), (21), (26) have been derived in a different way
in the papers [62-65]. In particular, in [64,65] it was shown that all the
above equations arise as the compatibility conditions for the system

\begin{equation}
\Delta _{i}\Phi =\psi T_{i}\psi ^{\ast },\quad i=1,2,3
\end{equation}
where $\Delta _{i}=T_{i}-1$ and $T_{i}$ are the Miwa shifts of the KP times,
i.e. $T_{i}\Phi (t)=\Phi (t+[a_{i}])=\Phi (t_{1}+a_{i},t_{2}+\frac{1}{2}%
a_{i}^{2},t_{3}+\frac{1}{3}a_{i}^{3},...).$

Equations (20), (21), (26) are closely connected with one more discrete
equation associated with the KP hierarchy, namely, with the famous bilinear
Hirota-Miwa equation

\begin{equation}
T_{1}\tau \cdot T_{2}T_{3}\tau -T_{2}\tau \cdot T_{3}T_{1}\tau +T_{3}\tau
\cdot T_{2}T_{1}\tau =0
\end{equation}
for the $\tau $-function. Solutions of equations (20),(21), (26) are ratios
of $\tau $-functions. The $\tau $- function is sort of homogeneous
coordinates for the lattice defined by these discrete equations.

\ One has discrete equations similar to equations (20), (21), (26), (28) for
multicomponent KP hierarchy, two-dimensional Toda lattice (2DTL) hierarchy
and other hierarchies [64,65]. \ They also have a nice geometrical
interpretation [49-51]. One of their common features is that all of them are
six and more points relations on the lattice.

\ The connection between BTs and DTs and discrete integrable equations helps
us also to clarify algebraic structures behind them. The IST linearizes not
only the nonlinear PDEs integrable by the method, but also the action of BTs
and DTs [66-68]. For instance, the action of the BT (7) which adds one
soliton to the solution of the KdV equation in term of the reflection
coefficient $R(\lambda )$ ( part of the inverse problem data) is a very
simple one

\begin{equation}
B_{\alpha }R(\lambda )=\frac{\lambda -i\alpha }{\lambda +i\alpha }R(\lambda
).
\end{equation}
Multiple action of BTs \ is then the multiplication by the rational function:

\begin{equation}
\prod_{k=1}^{n}B_{\alpha _{k}}\cdot R(\lambda )=\prod_{k=1}^{n}\frac{\lambda
-i\alpha _{k}}{\lambda +i\alpha _{k}}R(\lambda ).
\end{equation}

An action of BTs for other integrable equations has a similar form (see e.g.
[66-68]).

This property of BTs is inherited in the certain constructions of discrete
integrable equations [35-38,69,70]. For example, the method proposed in
[36-38] is based on the integral equation

\begin{equation}
\Psi (k)=\Psi _{0}(k)+\iint_{D}\Psi (l)d\mu (l,l^{\prime} )G(k,l^{\prime} )
\end{equation}
for the matrix-valued functions of the integers $n_{1},n_{2},n_{3}.$ Shift
in the variable $n_{i}$ is generated by the multiplication of the measure $%
d\mu (l,l^{\prime })$ by a simple rational function:

\begin{equation}
T_{i}:d\mu (l,l^{\prime })\rightarrow d\mu ^{\prime} (l,l^{\prime} )=\frac{%
l-p_{i}}{l^{\prime} +p_{i}}d\mu (l,l^{\prime })
\end{equation}
where $p_{i}$ are parameters.

Within the $\overline{\partial }$- dressing method based on the nonlocal $%
\overline{\partial }$ problem [71] (see also [72])

\begin{equation}
\frac{\partial \Psi (\lambda )}{\partial \overline{\lambda }}=\iint d\lambda
^{\prime} \Psi (\lambda ^{\prime} )R(\lambda ^{\prime} ,\lambda )
\end{equation}
an action of BT on the $\overline{\partial }$-data $R(\lambda \prime
,\lambda )$ is given by the formula [73,74]

\begin{equation}
B_{a}R(\lambda ^{\prime} ,\lambda )=\frac{\lambda ^{\prime} -a}{\lambda -a}%
R(\lambda ^{\prime} ,\lambda ).
\end{equation}

\ Both the $\overline{\partial }$- dressing method and the direct
linearization aprroach allow us to construct wide classes of discrete
integrable equations. In both these methods an action of multiple shifts on
the corresponding data is represented by the multiplication by rational
functions

\begin{equation}
\prod_{k=1}^{n}T_{\alpha _{k}}\cdot R(\lambda ^{\prime} ,\lambda
)=\prod_{k=1}^{n}\frac{\lambda ^{\prime} -a_{k}}{\lambda -a_{k}}R(\lambda
^{\prime} ,\lambda ).
\end{equation}

The method of constructing discrete equations proposed in [35] uses ,
essentially, the same idea.

\ Family of rational functions provides us with several examples of
associative algebras. One of them is the algebra of complex functions with
simple poles in distinct points. In virtue of the identity

\begin{equation}
\frac{a_{i}}{\lambda -\lambda _{i}}\cdot \frac{a_{k}}{\lambda -\lambda _{k}}%
=A_{i}\frac{a_{i}}{\lambda -\lambda _{i}}+A_{k}\frac{a_{k}}{\lambda -\lambda
_{k}},\quad i\neq k  \notag \\
\end{equation}
with $A_{i}=\frac{a_{k}}{\lambda _{i}-\lambda _{k}},A_{k}=\frac{a_{i}}{%
\lambda _{k}-\lambda _{i}}$ where $\lambda $ is a complex variable and $%
\lambda _{i},\lambda _{k},a_{i},a_{k}$ are arbitrary parameters the table of
multiplication for the elements $\mathbf{P}_{i}$ of the basis of this
algebra is of the form

\begin{equation}
\mathbf{P}_{i}\cdot \mathbf{P}_{k}=A_{i}\mathbf{P}_{i}+A_{k}\mathbf{P}%
_{k},\quad i\neq k,i,k=1,2,...,N.
\end{equation}

Functions $\mathbf{P}_{n}=\frac{a_{n}}{(\lambda -\lambda _{0})^{n}}$ with
multiple poles at the same point form an infinite-dimensional associative
algebra with the multiplication table $\mathbf{P}_{n}\cdot \mathbf{P}_{m}=%
\mathbf{P}_{n+m}$. A natural extension of this example to the polynomials $%
\mathbf{P}_{j}=\sum_{m=0}^{j}\frac{a_{jm}}{(\lambda -\lambda _{0})^{m}}$
provides us with the infinite-dimensional associative algebra with the
multiplication table

\begin{equation}
\mathbf{P}_{j}\cdot \mathbf{P}_{k}=\sum_{l=1}^{j+k}C_{jk}^{l}\mathbf{P}%
_{l},\quad j,k=1,2,...
\end{equation}
where $C_{jk}^{l}$ are certain constants ( structure constants). Under the
additional polynomial constraint $\mathbf{P}_{1}^{N+1}+u_{N}\mathbf{P}%
_{1}^{N}+...u_{1}\mathbf{P}_{1}=0$ the algebra (37) becomes the
N-dimensional associative algebra.

\ Associative algebras of the type (36) and (37) show up in the study of
many integrable systems both continuous and discrete. Within the methods
mentioned above the ring of rational functions and associative algebras did
not play a significant role. The relations of the type (36), (37) have
appeared only in certain intermediate calculations.

In a completely different contexts associative algebras have been used for
construction of integrable systems in the papers [75-80]. For instance, in
[80]they served basically to fix a domain of definition of dependent
variables for ODEs.

\ In the rest of the paper we shall try to demonstrate that the simple
associative algebras of the type (36), (37) are intimately connected with
integrable systems. The latter arise as the equations describing
deformations of the structure constants for such associative algebras.

\section{Deformations of associative algebras}

Here we will present basic elements of the approach to the deformations of
structure constants for associative algebras proposed in [21-24] in a
slightly modified form.

So , we consider a finite-dimensional noncommutative algebra \textit{A }with
( or without ) unite element $\mathbf{P}_{0}$. We will restrict overself to
a class of algebras which possess a basis composed by pairwise commuting
elements $\mathbf{P}_{0},\mathbf{P}_{1},...,\mathbf{P}_{N}$. The table of
multiplication

\begin{equation}
\mathbf{P}_{j}\cdot \mathbf{P}_{k}=\sum_{l=0}^{N}C_{jk}^{l}\mathbf{P}%
_{l},\quad j,k=0,1,...,N
\end{equation}
defines the structure constants $C_{jk}^{l}$ . The commutativity of the
basis implies that $C_{jk}^{l}$ $=C_{kj}^{l}$. In the presence of the unite
element one has $C_{j0}^{l}=\delta _{j}^{l}$ where $\delta _{j}^{l}$ is the
Kroneker symbol.

\ Following the Gerstenhaber's suggestion [3,4] we will treat the structure
constants $C_{jk}^{l}$ in a given basis as the objects to deform and will
denote the deformation parameters by $x^{1},x^{2},...,x^{M}$. \ In the
construction of deformations we should first to specify a ''deformed ''
version of the multiplication table (38) and then to require that this
realization is selfconsistence and meaningful.

\ Thus, to define deformations we

1) associate a set of elements $p_{0},p_{1},...,p_{N},x^{1},x^{2},...,x^{M}$
with the elements of the basis $\mathbf{P}_{0},\mathbf{P}_{1},...,\mathbf{P}%
_{N}$ and deformation parameters $x^{1},x^{2},...,x^{M}$,

2) consider the Lie algebra \textit{B} of the dimension N+M+1 with the basis
elements $e_{1},...,e_{N+M+1}$ obeying the commutation relations

\begin{equation}
\left[ e_{\alpha },e_{\beta }\right] =\sum_{\gamma =1}^{N+M+1}C_{\alpha
\beta \gamma }e_{\gamma },\quad \alpha ,\beta =1,2,...,N+M+1,
\end{equation}

3) identify the elements $p_{0},p_{1},...,p_{N},x^{1},x^{2},...,x^{M}$ with
the elements $e_{1},...,e_{N+M+1}$ thus defining the deformation driving
algebra (DDA). Different identifications define different DDAs. We will
assume that the element $p_{0}$ is always a central element of DDA. The
commutativity of the basis in the algebra \textit{A }implies the
commutativity between $p_{j}$ and in this paper we assume the same property
for all $x^{k}$. So, we will consider the DDAs defined by the commutation
relations of the type

\begin{equation}
\left[ p_{j},p_{k}\right] =0, \left[ x^{j},x^{k}\right] =0, \left[
p_{0},p_{k}\right] =0,\left[ p_{0},x^{k}\right] =0,\quad \left[ p_{j},x^{k}%
\right] =\sum_{l}\alpha _{jl}^{k}x^{l}+\sum_{l}\beta _{j}^{kl}p_{l}
\end{equation}
where $\alpha _{jl}^{k}$ and $\beta _{j}^{kl}$ are some constants,

4) consider the elements

\begin{equation}
f_{jk}=-p_{j}p_{k}+\sum_{l=0}^{N}C_{jk}^{l}(x)p_{l},\quad j,k=0,1,...,N
\end{equation}
of the universal enveloping algebra U(\textit{B}) of the algebra DDA(\textit{%
B}). These $f_{jk}$ ''represent'' the table (38) in U(\textit{B}),

5) require that the left ideal $J=\left\langle f_{jk}\right\rangle $
generated by these elements $f_{jk}$ is closed

\begin{equation}
\left[ J,J\right] \subset J
\end{equation}
or, equivalently , that

\begin{equation}
\left[ f_{jk},f_{lm}\right] =\sum_{s,t=0}^{N}\mathit{K}_{jklm}^{st}\cdot
f_{st},\quad j,k,l,m=0,1,...,N
\end{equation}
where $\mathit{K}_{jklm}^{st}$ are some elements of U(\textit{B}).

\textbf{Definition.} The structure constants $C_{jk}^{l}(x)$ are said to
define deformations of the algebra \textit{A} generated by given DDA if the
left ideal $J=\left\langle f_{jk}\right\rangle $ is closed.

\ To justify this definition we observe that the simplest possible
realization of the multiplication table (38) in U(\textit{B}) given by the
equations $f_{jk}=0$ is too restrictive. Indeed, the commutativity of $p_{j}$
implies in this case that $\left[ p_{t},C_{jk}^{l}(x)\right] =0$ and, hence,
no deformations are allowed. So, one should look for a weaker realization of
the multiplication table which is self-consistent. The condition that the
set of $f_{jk}$ form a closed ''algebra'' (43) is a natural candidate.

The condition (43) implies certain constraints on the structure constants.
The use of relations (40) provides us with the following identities

\begin{equation}
\left[ f_{jk},f_{lm}\right] =\sum_{s,t=0}^{N}\mathit{K}_{jklm}^{st}(x,p)%
\cdot f_{st}+\sum_{t=0}^{N}N_{jklm}^{t}(x)\cdot p_{t},\quad j,k,l,m=0,1,...,N
\end{equation}
and
\begin{equation}
(p_{j}p_{k})p_{l}-p_{l}(p_{k}p_{l})=\sum_{s,t=0}^{N}\mathit{L}%
_{klj}^{st}(x,p)\cdot f_{st}+\sum_{t=0}^{N}\Omega _{klj}^{t}(x)\cdot
p_{t},\quad j,k,l=0,1,...,N
\end{equation}
where $\mathit{K}_{jklm}^{st}(x,p),N_{jklm}^{s}(x),\mathit{L}%
_{klj}^{st}(x,p),\Omega _{klj}^{t}(x)$ are certain elements of U(\textit{B}%
). As an obvious consequence of the identity (44) one has

\textbf{Proposition 1. } Structure constants $C_{jk}^{l}(x)$ define
deformations generated by DDA if they obey the system of equations

\begin{equation}
N_{jklm}^{t}(x)=0,\quad j,k,l,m,s=0,1,...,N.\square
\end{equation}

Concrete form of $\mathit{K}_{jklm}^{st}(x,p),N_{jklm}^{t}(x),\mathit{L}%
_{klj}^{st}(x,p),\Omega _{klj}^{t}(x)$ and equations (46) is defined by the
DDA (40). In this paper we will consider as DDAs some three-dimensional Lie
algebras and algebras defined by the following commutation relations

\begin{equation}
\left[ p_{j},p_{k}\right] =0,\quad \left[ x^{j},x^{k}\right] =0,\quad \left[
p_{j},x^{k}\right] =\delta _{j}^{k}p_{j},\quad j,k=1,2,...,N
\end{equation}
and

\begin{equation}
\left[ p_{j},p_{k}\right] =0,\left[ x^{j},x^{k}\right] =0,\left[ p_{j},x^{k}%
\right] =\delta _{j}^{k}(p_{0}+\varepsilon _{j}p_{j}),\quad j,k=1,2,...,N
\end{equation}
where $\varepsilon _{j}$ are arbitrary parameters. In what follows we will
put the central element equal to the unite element $\widehat{I}$. Algebra of
shifts $p_{j}=T_{j}$ where $T_{j}x^{k}=x^{k}+\delta _{j}^{k},T_{j}\varphi
(x^{1},...,x^{N})=\varphi (x^{1},...,,x^{j}+1,...,x^{N})$ is a realization
of the algebra (47). A realization of the algebra (48) is given by the
algebra of differences $p_{j}=\Delta _{j}=\frac{T_{j}-\widehat{I}}{%
\varepsilon _{j}}$ where $T_{j}x^{k}=x^{k}+\varepsilon _{j}\delta _{j}^{k}$.
The family of algebras (48) contains the Heisenberg algebra as the limit
when all $\varepsilon _{j}\rightarrow 0$. In this case $\Delta
_{j}\rightarrow \frac{\partial }{\partial x^{j}}$. At $\varepsilon
_{j}=1(j=1,...,N)$ one has the algebra connected with the algebra (47) by
the change of the basis $\mathbf{P}_{j}\longleftrightarrow \mathbf{P}_{j}+%
\mathbf{P}_{0}$ in the algebra \textit{A}.

In order to calculate explicitly the r.h.s. in the identities (44) and (45)
one needs to know the commutator $\left[ p_{t},C_{jk}^{l}(x)\right] $. \ For
the algebra (47) ( i.e. DDA (47)) and an element $\varphi (x^{1},...,x^{N})$
$\subset U(DDA(47))$ one has

\begin{equation}
\left[ p_{j},\varphi (x)\right] =\Delta _{j}\varphi (x)\cdot p_{j},\quad
j=1,...,N
\end{equation}
where $\Delta _{j}=T_{j}-1$ and $T_{j}$ is the shift operator $%
T_{j}x^{k}=x^{k}+\delta _{j}^{k}$. For the DDA (48) the analogous identity is

\begin{equation}
\left[ p_{j},\varphi (x)\right] =\Delta _{j}\varphi (x)\cdot (\widehat{I}%
+\varepsilon _{j}p_{j}),\quad j=1,...,N
\end{equation}
where $\Delta _{j}=\frac{T_{j}-\widehat{I}}{\varepsilon _{j}}$ and $%
T_{j}x^{k}=x^{k}+\varepsilon _{j}\delta _{j}^{k}$.

Using (49), for the DDA (47) one gets

\begin{align}
\mathit{K}_{jklm}^{st}=& \frac{1}{2}\Bigl( \delta _{j}^{s}\delta
_{k}^{t}%
\sum_{n}(T_{j}T_{k}C_{lm}^{n})p_{n}+(T_{j}T_{k}C_{lm}^{s})(T_{s}C_{jk}^{t})+
\notag \\
& \qquad + \delta _{l}^{s}\delta _{m}^{t}
\sum_{n}C_{jk}^{n}p_{n}-(j,l)(k,m)+(s,t)\Bigr)  \notag
\end{align}
where the bracket (j,k) denote the previous terms with the exchange of
indices indicated in the bracket,

\begin{equation}
N_{jklm}^{t}=\sum_{n,s}(T_{l}T_{m}C_{jk}^{s})(T_{s}C_{mn}^{n})C_{ns}^{t}-%
\sum_{n,s}(T_{j}T_{k}C_{lm}^{s})(T_{s}C_{jk}^{n})C_{ns}^{t}
\end{equation}
and

\begin{equation}
\mathit{L}_{klj}^{st}=\frac{1}{2}\left( \delta _{k}^{s}\delta
_{l}^{t}p_{j}-\delta _{j}^{s}\delta _{k}^{t}p_{l}+\delta
_{j}^{s}(T_{j}C_{kl}^{t})-\delta _{l}^{s}(T_{l}C_{jk}^{t})+(s,t)\right) ,
\end{equation}

\begin{equation}
\Omega
_{klj}^{t}=\sum_{s}(C_{ls}^{t}(T_{l}C_{jk}^{s})-C_{sj}^{t}(T_{j}C_{kl}^{s})).
\end{equation}

For the DDA (48) the use of (50) gives

\begin{align}
\mathit{K}_{jklm}^{st}= \frac{1}{2}\Bigl( \delta _{j}^{s}\delta
_{k}^{t}\sum_{n}(T_{j}T_{k}C_{lm}^{n}) p_{n}+
(T_{j}T_{k}C_{lm}^{s})(T_{s}C_{jk}^{t})+  \notag \\
\delta _{j}^{t}(T_{j}\Delta _{k}C_{lm}^{s})+\delta _{k}^{t}(T_{k}\Delta
_{j}C_{lm}^{s})-(j, l)(k, m)+(s,t)\Bigr) ,  \notag
\end{align}

\begin{align}
N_{jklm}^{t}=-\Delta _{j}\Delta _{k}C_{lm}^{t}-\sum_{s}\Bigl(
C_{js}^{t}T_{j}\Delta _{k}C_{lm}^{s}+C_{ks}^{t}T_{k}\Delta _{j}C_{lm}^{s}+
\notag \\
+ (T_{j}T_{k}C_{lm}^{s})(\Delta _{s}C_{jk}^{t})+
\sum_{n}(T_{j}T_{k}C_{lm}^{s})(T_{s}C_{jk}^{n})C_{ns}^{t}\Bigr) -(j, l)(k, m)
\end{align}
and

\begin{equation}
\mathit{L}_{klj}^{st}=\frac{1}{2}\left( \delta _{k}^{s}\delta
_{l}^{t}p_{j}-\delta _{j}^{s}\delta _{k}^{t} p_{l}+\delta _{j}^{s}(\Delta
_{j}C_{kl}^{t})-\delta _{l}^{s}(\Lambda _{l}C_{jk}^{t})+(s, t)\right) ,
\end{equation}

\begin{equation}
\Omega _{klj}^{t}=\Delta _{l}C_{jk}^{t}-\Delta
_{j}C_{lk}^{t}+%
\sum_{s}(C_{ls}^{t}(T_{l}C_{jk}^{s})-C_{sj}^{t}(T_{j}C_{kl}^{s})).
\end{equation}

Thus, deformations generated by DDAs (47) and (48) are governed by equations
(46) with $N_{jklm}^{t}$ given by (52) and (53), respectively.

\section{Class of integrable deformations}

\ The l.h.s. of equation (46) has a special structure in both cases under
consideration. Indeed, one can show that for the DDA (47)

\begin{align}
\ N_{jklm}^{t}=\frac{1}{2}\sum_{s}\Bigl( C_{ls}^{t}(T_{l}\Omega
_{kmj}^{s})-C_{js}^{t}(T_{j}\Omega _{mkl}^{s})-(T_{j}T_{l}C_{km}^{s})\Omega
_{sjl}^{t}+  \notag \\
+(T_{j}T_{k}C_{lm}^{s})\Omega_{ksj}^{t}+(T_{m}T_{l}C_{jk}^{s})\Omega
_{msl}^{t}+(j, k)(l,m)\Bigr)
\end{align}
while for DDA (48) one has

\begin{align}
N_{jklm}^{t}=\Delta _{m}\Omega _{jlk}^{t}-\Delta _{k}\Omega _{ljm}^{t}+\frac{%
1}{2}\sum_{s}\Bigl( C_{ks}^{t}(T_{k}\Omega
_{lmj}^{s})+C_{ms}^{t}(T_{m}\Omega _{jlk}^{s})+ (T_{j}T_{k}C_{lm}^{s})\Omega
_{jks}^{t}+  \notag \\
+(T_{l}T_{m}C_{jk}^{s})\Omega _{lsm}^{t}+(T_{k}T_{m}C_{jl}^{s})\Omega
_{smk}^{t}+(j,k)(l, m)\Bigr) .
\end{align}

The r.h.s. of these formulae vanish if the structure constants obey the
equation $\Omega _{klj}^{t}=0$. Thus, one has

\textbf{Proposition 2. }\ The equations

\begin{equation}
\sum_{s}(C_{ls}^{t}(T_{l}C_{jk}^{s})-C_{sj}^{t}(T_{j}C_{kl}^{s}))=0
\end{equation}
and

\begin{equation}
\Delta _{l}C_{jk}^{t}-\Delta
_{j}C_{lk}^{t}+%
\sum_{s}(C_{ls}^{t}(T_{l}C_{jk}^{s})-C_{sj}^{t}(T_{j}C_{kl}^{s}))=0
\end{equation}
govern the subclasses of deformations generated by the DDA (47) and DDA
(48), respectively.

\ In the rest of the paper we will study only these classes of deformations.
We will refer to the systems of equations (59), (60) as the central systems
(CSs). There are at least two reasons to consider these subclasses of
deformations. The first is that they are equivalent to the compatibility
conditions of the linear systems

\begin{equation}
f_{jk}\Psi =0,\quad j,k,=1,...,N
\end{equation}
where $\Psi $ is a common right divisor of zero for all $f_{jk}$. Recall
that non-zero elements a and b of an algebra are called left and right
divisors of zero if ab=0 (see e.g.[81]). Within the interpretation of $p_{j}
$ and $x^{k}$ as operators acting in a linear space \textit{H} equations
(61) become the following linear problems for integrable systems

\begin{equation}
\left( -p_{j}p_{k}+\sum_{l}C_{jk}^{l}(x)p_{l}\right) \left| \Psi
\right\rangle =0,\quad j,k=1,...,N
\end{equation}
where $\left| \Psi \right\rangle \subset \mathit{H}$. So, one can refer to
such deformations as integrable one.

\ The second reason to study equations (59), (60) is that they have a nice
geometrical meaning. We begin this study with rewriting these equations in a
compact form. Introducing the operators $T_{j}^{C}$ and $\nabla _{j}$ acting
as

\begin{equation}
T_{j}^{C}\Phi _{k}^{n}= \sum_{s}C_{js}^{n}T_{j}\Phi _{k}^{s}
\end{equation}
and

\begin{equation}
\nabla _{j}\Phi _{k}^{n}=\Delta _{j}\Phi
_{k}^{n}+\sum_{s}C_{js}^{n}T_{j}\Phi _{k}^{s},
\end{equation}
one gets the following form of the formulae (51) and (54)

\begin{equation}
N_{jklm}^{t}=%
\sum_{n}((T_{l}T_{m}C_{jk}^{n})T_{n}^{C}C_{lm}^{t}-(T_{j}T_{k}C_{lm}^{n})T_{n}^{C}C_{jk}^{t})
\end{equation}
and

\begin{eqnarray}
N_{jklm}^{t}=\nabla _{l}\nabla _{m}C_{jk}^{t}-\nabla _{j}\nabla
_{k}C_{lm}^{t}+\sum_{n}((T_{l}T_{m}C_{jk}^{n})(\nabla _{l}C_{nm}^{t}-\nabla
_{m}C_{nl}^{t})-  \notag \\
\sum_{n}((T_{j}T_{k}C_{lm}^{n})(\nabla _{j}C_{nk}^{t}-\nabla _{k}C_{nj}^{t})
\end{eqnarray}
respectively. For $\Omega _{klj}^{t}$ (53), (56) one gets

\begin{equation}
\Omega _{klj}^{t}=T_{l}^{C}C_{jk}^{t}-T_{j}^{C}C_{lk}^{t}
\end{equation}
and

\begin{equation}
\Omega _{klj}^{t}=\nabla _{l}C_{jk}^{t}-\nabla _{j}C_{lk}^{t}.
\end{equation}

Then, introducing the matrices $C_{j}$ and $\Omega _{lj}$ such that $%
(C_{j})_{k}^{l}=C_{jk}^{l}$ and ($\Omega _{lj})_{k}^{t}=\Omega _{klj}^{t}$ ,
one rewrites equations (59) and (60) in the matrix form

\begin{equation}
\Omega_{lj}=C_{l}T_{l}C_{j}-C_{j}T_{j}C_{l}=T_{l}^{C}C_{j}-T_{j}^{C}C_{l}=0
\end{equation}
and

\begin{equation}
\Omega _{lj}=\Delta _{l}C_{j}-\Delta
_{j}C_{l}+C_{l}T_{l}C_{j}-C_{j}T_{j}C_{l}=\nabla _{l}C_{j}-\nabla
_{j}C_{l}=0.
\end{equation}

Note that equation (70) is equivalent to the equation

\begin{equation}
(1+\varepsilon _{l}C_{l})T_{l}(1+\varepsilon _{j}C_{j})-(1+\varepsilon
_{j}C_{j})T_{j}(1+\varepsilon _{l}C_{l})=0
\end{equation}
which is of the form (69) for the matrix $\widetilde{C}_{j}=1+\varepsilon
_{j}C_{j}$. One observes this similarity also for the operators $T_{j}^{C}$
and $\nabla _{j}$ which in the matrix notations act as follows

\begin{equation}
T_{j}^{C}=C_{j}T_{j},\nabla _{j}=\Delta _{j}+C_{j}T_{j}=\frac{1}{\varepsilon
_{j}}\left( (1+\varepsilon _{j}C_{j})T_{j}-1\right) .
\end{equation}

\ For constant structure constants CSs (69), (60) are reduced to the
associativity condition (3), i.e. $\left[ C_{l},C_{j}\right] =0$. In
general, for deformed $C_{jk}^{l}(x)$ this condition is not satisfied. The
defect of associativity or quantum anomaly for deformations [21] is defined
as the matrix $\alpha _{lj}=\Omega _{lj}-$ $\left[ C_{l},C_{j}\right] $ .
For deformations generated by DDA (47) it is equal to

\begin{equation}
\alpha _{lj}=C_{l}\Delta _{l}C_{j}-C_{j}\Delta _{j}C_{l}
\end{equation}
while for DDA (48)

\begin{equation}
\alpha _{lj}=(1+\varepsilon _{l}C_{l})\Delta _{l}C_{j}-(1+\varepsilon
_{j}C_{j})\Delta _{j}C_{l}.
\end{equation}

\ To clarify the geometrical content of equation (70) we note that in the
case of all $\varepsilon _{j}=0$ it is reduced to that of quantum
deformations [21], i.e. to the system

\begin{equation}
\Omega _{klj}^{t}=\frac{\partial C_{kj}^{t}}{\partial x^{l}}-\frac{\partial
C_{kl}^{t}}{\partial x^{j}}%
+\sum_{m}(C_{jk}^{m}C_{lm}^{t}-C_{lk}^{m}C_{jm}^{t})=0,\quad j,k,l,t=1,...,N.
\end{equation}
This equation has a geometrical meaning of vanishing Riemann curvature
tensor $(R_{lj}^{class})_{k}^{t}\doteqdot R_{klj}^{t}=\Omega _{klj}^{t}$
with the Christoffel symbols identified with the structure constants $%
C_{jk}^{l}$ [21]. Operator $\nabla _{j}$ becomes a covariant derivative $%
\nabla _{j}=\frac{\partial }{\partial x^{j}}+C_{j}$ and one has (see e.g.
[82])

\begin{equation}
R_{jk}^{class}=\left[ \nabla _{j},\nabla _{k}\right] .
\end{equation}
In particular, the equation $R_{lj}^{class}=0$ is equivalent to the
compatibility condition for the linear problems

\begin{equation}
\nabla _{j}\Psi =0,\quad j=1,...,N.
\end{equation}

For a general DDA (48) one observes that

\begin{equation}
\left[ \nabla _{j},\nabla _{k}\right] =\Omega _{jk}T_{j}T_{k}.
\end{equation}
By analogy with (76) equation (78) can be understood as the definition of
the discrete version $R_{jk}^{d}$ of the curvature tensor $R_{jk}^{class}$:

\begin{equation}
\left[ \nabla _{j},\nabla _{k}\right] =R_{jk}^{d}T_{j}T_{k}.
\end{equation}
Thus, $R_{jk}^{d}=\Omega _{jk}$ or in components

\begin{equation}
R_{klj}^{dt}=\Omega _{klj}^{t}=\Delta _{l}C_{jk}^{t}-\Delta
_{j}C_{lk}^{t}+%
\sum_{s}(C_{ls}^{t}(T_{l}C_{jk}^{s})-C_{sj}^{t}(T_{j}C_{kl}^{s})).
\end{equation}
Obviously, $\lim_{\varepsilon _{j}\rightarrow
0}R_{klj}^{dt}=R_{klj}^{(class)t}.$

Similar to the continuous case the CS (70) is equivalent to the equation $%
\left[ \nabla _{l},\nabla _{j}\right] =0$ and to the compatibilty condition
for the linear problems

\begin{equation}
\nabla _{j}\Psi =(\Delta _{j}+C_{j}T_{j})\Psi =0,\quad j=1,...,N.
\end{equation}

\ Amazingly, the ''tensor'' (80) and the operator $\nabla _{j}$ (64)
essentially coincide with the discrete Riemann curvature tensor and
covariant derivative introduced earlier within various discretizations of
Riemann geometry ( see e.g. [83-85]). Thus, the CS which governs the
deformations of the structure constants $C_{jk}^{l}(x)$ have the same
geometrical meaning of vanishing Riemann curvature tensor both in continuous
and difference cases.

Finally, we note that the CSs (59), (60) or (69), (70) are underdetermined
systems of equations. Similar to the integrable PDEs (see e.g. [25-28]) it
is connected with the gauge freedom. It is not difficult to see that, for
instance, equation (69) is invariant under transformations

\begin{equation}
C_{j}\rightarrow \widetilde{C}_{j}=GC_{j}T_{j}G^{-1},
\end{equation}
where $G(x)$ is a diagonal matrix with the diagonal elements $G_{k}(x)$ $%
\subset U(DDA)$ generated only by the elements $x^{1},...,x^{N}$, since
under this transformation

\begin{equation}
\widetilde{\Omega }_{lj}\doteqdot \widetilde{C}_{l}T_{l}\widetilde{C}_{j}-%
\widetilde{C}_{j}T_{j}\widetilde{C}_{l}=G\Omega _{lj}T_{l}T_{j}G^{-1}.
\end{equation}

The relation $C_{jk}^{l}=C_{kj}^{l}$ implies that $G_{k}=T_{k}g(x)$ where
g(x) is an arbitrary element of $U(DDA)$ generated by $x^{1},...,x^{N}$. So,
the CS (69) is invariant under the transformations

\begin{equation}
C_{jk}^{l}\rightarrow \widetilde{C}_{jk}^{l}=T_{l}g\cdot
(T_{j}T_{k}g^{-1})C_{jk}^{l}
\end{equation}
and for the elements $f_{jk}$ one has

\begin{equation}
f_{jk}\rightarrow \widetilde{f}_{jk}\doteqdot -p_{j}p_{k}+\sum_{l}\widetilde{%
C}_{jk}^{l}(x)p_{l}=(T_{j}T_{k}g^{-1})\cdot f_{jk}\cdot g.
\end{equation}
Analogously, the CS (70) is invariant under the transformations

\begin{equation}
C_{j}\rightarrow \widetilde{C}_{j}=G\Delta _{j}G^{-1}+GC_{j}T_{j}G^{-1}.
\end{equation}

\ In the continuous case ($\varepsilon _{j}\rightarrow 0,T_{j}\rightarrow
1,\Delta _{j}\rightarrow \frac{\partial }{\partial x^{j}}$) the
transformations (86) are well-known in the theory of integrable equations as
the gauge transformations. Transformations (84), (86) are their discrete and
difference counterparts (see also [83-85]). The invariance of the CSs under
these transformations means that deformations governed by them form the
classes of gauge equivalent deformations. Note that the associativity
conditions (3) themselves are not invariant under gauge transformations (84).

\section{ Discrete deformations of three-dimensional algebra and Boussinesq
equation.}

\ A simplest nontrivial example of the proposed scheme corresponds to the
three-dimensional algebra with the unite element and the basis $\mathbf{P}%
_{0},\mathbf{P}_{1},\mathbf{P}_{2}.$ The table of multiplication is given by
the trivial part $\mathbf{P}_{0}\mathbf{P}_{j}=\mathbf{P}_{j},j=0,1,2$ and by

\begin{eqnarray}
\mathbf{P}_{1}^{2}=A\mathbf{P}_{0}+B\mathbf{P}_{1}+C\mathbf{P}_{2},  \notag
\\
\mathbf{P}_{1}\mathbf{P}_{2}=D\mathbf{P}_{0}+E\mathbf{P}_{1}+G\mathbf{P}_{2},
\notag \\
\mathbf{P}_{2}^{2}=L\mathbf{P}_{0}+M\mathbf{P}_{1}+N\mathbf{P}_{2}
\end{eqnarray}%
where the structure constants A,B,...,N depend only on the deformation
parameters $x^{1},x^{2}$.  It is convenient also to arrange the structure
constants A,B,...,N into the matrices $C_{1},C_{2}$ defined as above by $%
(C_{j})_{k}^{l}=C_{jk}^{l}$. One has

\begin{equation}
C_{1}=\left(
\begin{array}{ccc}
0 & A & D \\
1 & B & E \\
0 & C & G%
\end{array}%
\right) ,\quad C_{2}=\left(
\begin{array}{ccc}
0 & D & L \\
0 & E & M \\
1 & G & N%
\end{array}%
\right) .
\end{equation}
In terms of these matrices the associativity conditions (2) are written as

\begin{equation}
C_{1}C_{2}=C_{2}C_{1}.
\end{equation}
and the CSs (69) and (70) are

\begin{equation}
C_{1}T_{1}C_{2}=C_{2}T_{2}C_{1}
\end{equation}
and

\begin{equation}
\Delta _{1}C_{2}-\Delta _{2}C_{1}+C_{1}T_{1}C_{2}-C_{2}T_{2}C_{1}=0
\end{equation}
respectively.

Let us consider first the CS (91). In terms of A, B,... it is the system

\begin{eqnarray}
\Delta _{1}D-\Delta _{2}A+AE_{1}+DG_{1}-DB_{2}-LC_{2} &=&0,  \notag \\
\Delta _{1}E-\Delta _{2}B+D_{1}+BE_{1}+EG_{1}-EB_{2}-MC_{2} &=&0,  \notag \\
\Delta _{1}G-\Delta _{2}C+CE_{1}+GG_{1}-A_{2}-GB_{2}-NC_{2} &=&0,  \notag \\
\Delta _{1}L-\Delta _{2}D+AM_{1}+DN_{1}-DE_{2}-LG_{2} &=&0,  \notag \\
\Delta _{1}M-\Delta _{2}E+L_{1}+BM_{1}+EN_{1}-EE_{2}-MG_{2} &=&0,  \notag \\
\Delta _{1}N-\Delta _{2}G+CM_{1}+GN_{1}-D_{2}-GE_{2}-NG_{2} &=&0
\end{eqnarray}%
where $A_{j}\doteqdot T_{j}A$ etc. \ As in the continuous case [21] we
consider the \ constraint B=0, C=1, G=0. The first three equations (92) give

\begin{equation}
L=\Delta _{1}D-\Delta _{2}A+AE_{1},\quad M=\Delta _{1}E+D_{1},\quad
N=E_{1}-A_{2}
\end{equation}
and the rest of the system (92) takes the form

\begin{eqnarray}
\left( \Delta _{1}^{2}-\Delta _{2}+E_{11}-E_{2}-A_{12}\right) E+2\Delta
_{1}D_{1}-\Delta _{2}A_{1}+A_{1}E_{11} =0,  \notag \\
\left( \Delta _{1}^{2}-\Delta _{2}+E_{11}-E_{2}-A_{12}\right) D+\left(
-\Delta _{1}\Delta _{2}+\Delta _{1}E_{1}+D_{11}\right) A+\Delta _{1}(AE_{1})
=0,  \notag \\
\Delta _{1}(2E_{1}-A_{2})+D_{11}-D_{2} =0.
\end{eqnarray}

In the continuous case $\varepsilon _{1}=\varepsilon _{2}=0$, i.e. for the
Heisenberg DDA the CS (94) becomes ( $A_{x_{j}}=\frac{\partial A}{\partial
x^{j}},$ etc)

\begin{eqnarray}
E_{x_{1}x_{1}}-E_{x_{2}}+2D_{x_{1}}-A_{x_{2}} =0,  \notag \\
D_{x_{1}x_{1}}-D_{x_{2}}-A_{x_{1}x_{2}}+AE_{x_{1}}+(AE)_{x_{1}} =0,\quad
2E-A=0
\end{eqnarray}%
where all integration constants have been choosen to be equal to zero. Hence
, one has the system

\begin{equation}
\frac{1}{2}A_{x_{1}x_{1}}-\frac{3}{2}A_{x_{2}}+2D_{x_{1}}=0,\quad
D_{x_{1}x_{1}}-D_{x_{2}}-A_{x_{1}x_{2}}+\frac{3}{4}(A^{2})_{x_{1}}=0.
\end{equation}
Eliminating D, one gets the Boussinesq (BSQ) equation

\begin{equation}
A_{x_{2}x_{2}}+\frac{1}{3}A_{x_{1}x_{1}x_{1}x_{1}}-(A^{2})_{x_{1}x_{1}}=0.
\end{equation}

The BSQ equation (97) defines quantum deformations of the structure
constants in (87) with $B=0,C=1,G=0,L=D_{x_{1}}-A_{x_{2}}+\frac{1}{2}A^{2},M=%
\frac{1}{2}A_{x_{1}}+D,N=-\frac{1}{2}A$ and A, D defined by (96).

In the pure discrete case $\varepsilon _{1}=\varepsilon _{2}=1$ the CS (94)
represents the discrete version of the BSQ system (96) which is equivalent
to that proposed in [37]. It defines the discrete deformations of the same
structure constants.

There are also the mixed cases. The first is $\varepsilon _{1}=0,\varepsilon
_{2}=1(T_{1}=1,\Delta _{1}=\frac{\partial }{\partial x^{1}},\Delta
_{2}=T_{2}-1)$ . The CS (94) is

\begin{eqnarray}
E_{x_{1}x_{1}}+2D_{x_{1}}-(1+E)\Delta _{2}(E+A) =0,  \notag \\
D_{x_{1}x_{1}}-D_{x_{2}}+AE_{x_{1}}+(AE)_{x_{1}}-\Delta
_{2}A_{x_{1}}-D\Delta _{2}(E+A) =0, \\
(2E-A_{2})_{x_{1}}-\Delta _{2}D_{2} =0.  \notag
\end{eqnarray}

The second case corresponds to $\varepsilon _{1}=1,\varepsilon _{2}=0(\Delta
_{1}=T_{1}-1,T_{2}=1,\Delta _{2}=\frac{\partial }{\partial x^{2}})$ and CS
takes the form

\begin{eqnarray}
\Delta _{1}^{2}E+(E_{11}-E-A_{1})E+A_{1}E_{11}+2\Delta
_{1}D_{1}-(E+A_{1})_{x_{2}} =0,  \notag \\
\Delta _{1}^{2}D+(E_{11}-E-A_{1})D+A\Delta _{1}E_{1}+\Delta
_{1}(AE_{1})+D_{11}A-\Delta _{1}A_{x_{2}}-D_{x_{2}} =0,  \notag \\
\Delta _{1}(2E_{1}-A)+D_{11}-D =0.
\end{eqnarray}

Equations (62) provide us with the linear problems for the CS (94). They are

\begin{equation}
\left( \Delta _{2}-\Delta _{1}^{2}+A\right) \mid \Psi \rangle =0,\quad
\left( \Delta _{1}\Delta _{2}-E\Delta _{1}-D\right) \mid \Psi \rangle =0
\end{equation}
or equivalently

\begin{equation}
\left( \Delta _{2}-\Delta _{1}^{2}+A\right) \mid \Psi \rangle =0,\quad
\left( \Delta _{1}^{3}-(E+A)\Delta _{1}-(D+\Delta _{1}A)\right) \mid \Psi
\rangle =0.
\end{equation}
One can check that the equation $f_{22}\mid \Psi \rangle =0$ is a
consequence of (100).

In the quantum case ($\Delta _{j}=\frac{\partial }{\partial x^{j}})$ the
system (101) is the well-known linear system for the continuous BSQ equation
[86]. We note that in this case the third equation $f_{22}\mid \Psi \rangle
=0$ , i.e. $\left( \Delta _{2}^{2}-L-D\Delta _{1}+\frac{1}{2}A\right) \mid
\Psi \rangle =0$ is the consequence of two equations (101).

In the pure discrete case equations (101) represent the linear problems for
the discrete BSQ equation. The first mixed case considered above can be
treated as the BSQ equation with the discrete time. The second case instead
represents a continuous isospectral flow for the third order difference
problem, i.e. a sort of the difference BSQ equation with continuous time.

We would like to emphasize that equations (96),(99) and (92) describe
different classes of deformations of the same set of the structure constants
A,B, ...,N defined by the table (87).

\ Now let us consider the CS (90). It has the form (92) where the terms with
the differences $\Delta _{j}A$ etc should be dropped out. In the BSQ gauge
B=0, C=1, G=0 it becomes

\begin{eqnarray}
(E_{11}-E_{2}-A_{12})E+A_{1}E_{11}=0,  \notag \\
(E_{11}-E_{2}-A_{12})D+A_{1}D_{11}=0,D_{11}-D_{2}=0.
\end{eqnarray}
Note that this system imlies that $A_{1}E_{11}D-AED_{11}=0$. With the choice
D=1 the system (102) takes the form

\begin{equation}
AE=(AE_{1})_{1},\quad A_{12}-A=E_{11}-E_{2}.
\end{equation}
Introducing the function U such that $A=U_{11}-U_{2},E=U_{12}-U$, one
rewrites the system (103) as the single equation

\begin{equation}
(U_{12}-U)(U_{11}-U_{2})=\left( (U_{112}-U_{1})(U_{11}-U_{2})\right) _{1}.
\notag \\
\end{equation}
This discrete equation governs deformations of the BSQ structure constants
generated by DDA (47).

\section{ WDVV, discrete and continuous-discrete WDVV equations}

Another interesting reduction of the CS (92) corresponds to the constraint $%
C=1,G=0,N=0.$ In this case the CS is

\begin{eqnarray}
\Delta _{1}D-\Delta _{2}A+AE_{1}-DB_{2}-L &=&0,  \notag \\
\Delta _{1}E-\Delta _{2}B+D_{1}+BE_{1}-EB_{2}-M &=&0,  \notag \\
\Delta _{1}L-\Delta _{2}D+AM_{1}-DE_{2} &=&0,  \notag \\
\Delta _{1}M-\Delta _{2}E+L_{1}+BM_{1}-EE_{2} &=&0,  \notag \\
E_{1}-A_{2} &=&0,M_{1}-D_{2}=0.
\end{eqnarray}
Third and sixth equations (104) imply the existence of the functions U and V
such that
\begin{equation}
A=U_{1},\quad E=U_{2},\quad D=V_{1},\quad M=V_{2}.
\end{equation}
Substituting the expressions for L and M given by the first two equations
(104), i.e.

\begin{equation}
L =\Delta _{1}V_{1}-\Delta _{2}U_{1}+U_{1}U_{12}-V_{1}B_{2},\quad M =\Delta
_{1}U_{2}-\Delta _{2}B+V_{11}+BU_{12}-U_{2}B_{2}  \notag \\
\end{equation}
into the rest of the system , one gets the following three equations

\begin{eqnarray}
\Delta _{1}U_{1}-\Delta _{2}B+V_{11}+BU_{12}-U_{2}B_{2}-V_{2} =0, \\
(\Delta _{1}^{2}-\Delta _{2})V_{1}-\Delta _{1}\Delta _{2}U_{1}+\Delta
_{1}(U_{1}U_{12})-\Delta _{1}(V_{1}B_{2})-V_{1}U_{22}+U_{1}V_{12} =0,  \notag
\\
\Delta _{1}(V_{2}+V_{11})-\Delta
_{2}(U_{2}+U_{11})+U_{11}U_{112}-U_{2}U_{22}-V_{11}B_{12}+BV_{12} =0.  \notag
\end{eqnarray}
Solution of this system together with the formulas (105) define deformations
of the structure constants A,B,... generated by the DDA (48).

In the pure continuous case ( $\varepsilon _{1}=\varepsilon _{2}=0$ ) the
system (106) takes the form

\begin{eqnarray}
U_{x_{1}}-B_{x_{2}} &=&0,\quad V_{x_{1}}-U_{x_{2}}=0,  \notag \\
(V_{x_{1}}+U_{x_{2}}+U^{2}-VB)_{x_{1}}-V_{x_{2}} &=&0.
\end{eqnarray}
This system of three conservation laws implies the existence of the function
F such that

\begin{equation}
U=F_{x_{1}x_{1}x_{2}},\quad V=F_{x_{1}x_{2}x_{2}}, \quad
B=F_{x_{1}x_{1}x_{1}}.
\end{equation}
In terms of F the system (107) is

\begin{equation}
F_{x_{2}x_{2}x_{2}}-(F_{x_{1}x_{1}x_{2}})^{2}+F_{x_{1}x_{1}x_{1}}F_{x_{1}x_{2}x_{2}}=0.
\end{equation}
It is the WDVV\ equation (1) with the metric $g=\left(
\begin{array}{ccc}
0 & 0 & 1 \\
0 & 1 & 0 \\
1 & 0 & 0%
\end{array}%
\right) $ [10,11]. It is a well-known fact the WDVV equation describes
deformations of the three-dimensional algebra (87) under the reduction $%
C=1,G=0,N=0$ [3-10]. So, the system (106) represents the generalization of
the WDVV equation to the case of deformations of the same algebra generated
by DDA (48).

In the pure discrete case $\varepsilon _{1}=\varepsilon _{2}=1$ the system
(106) gives us a pure discrete version of the WDVV equation. In the first
mixed case $\varepsilon _{1}=0,\varepsilon _{2}=1$ we have the system

\begin{eqnarray}
U_{x_{1}}-\Delta _{2}(B+V)-U_{2}\Delta _{2}B =0, \\
(V_{x_{1}}+UU_{2}-VB_{2})_{x_{1}}-\Delta _{2}(V+U_{x_{1}})-VU_{22}+UV_{2} =0,
\notag \\
(V+V_{2})_{x_{1}}-\Delta _{2}(U+U_{2}+UU_{2})-VB_{2}+BV_{2} =0  \notag
\end{eqnarray}
while in the case $\varepsilon _{1}=1,\varepsilon _{2}=0$ one gets

\begin{eqnarray}
\Delta _{1}U_{1}+B\Delta _{1}U+V_{11}-V-B_{x_{2}} =0, \\
\Delta _{1}(\Delta
_{1}V_{1}-U_{1x_{2}}+U_{1}^{2}-V_{1}B)-V_{1}U+U_{1}V_{1}-V_{1x_{2}} =0,
\notag \\
\Delta _{1}(V+V_{11})+U_{11}^{2}-U^{2}-V_{11}B_{1}+BV_{1}-(U+U_{11})_{x_{2}}
=0.  \notag
\end{eqnarray}

Equations $f_{jk}\mid \Psi \rangle =0$ for the system (104) in the
coordinate representation $p_{j}=\Delta _{j}$ have the form

\begin{eqnarray}
\left( \Delta _{2}-\Delta _{1}^{2}+B\Delta _{1}+A\right) \mid \Psi \rangle
=0,\quad \left( \Delta _{1}\Delta _{2}-E\Delta _{1}-D\right) \mid \Psi
\rangle =0,  \notag \\
\left( \Delta _{2}^{2}-M\Delta _{1}-L\right) \mid \Psi \rangle =0.
\end{eqnarray}

The system (112) in its turn is equivalent to the following

\begin{eqnarray}
\left( \Delta _{1}^{3}-B\Delta _{1}^{2}-(\Delta _{1}B+A_{1}+E)\Delta
_{1}-(\Delta _{1}A+D)\right) \mid \Psi \rangle =0,  \notag \\
\left( \Delta _{2}-\Delta _{1}^{2}+B\Delta _{1}+A\right) \mid \Psi \rangle
=0,\quad \left( \Delta _{2}^{2}-M\Delta _{1}-L\right) \mid \Psi \rangle =0.
\end{eqnarray}

The compatibility condition for the above linear problems are equivalent to
the ''discretized'' WDVV systems (104) or (106). In particular, in the pure
continuous case the problems (112) in the coordinate representation coincide
with the well-known one [12,13] , namely,

\begin{eqnarray}
\Psi _{x_{1}x_{1}} &=&F_{x_{1}x_{1}x_{2}}\Psi +F_{x_{1}x_{1}x_{1}}\Psi
_{x_{1}}+\Psi _{x_{2}},  \notag \\
\Psi _{x_{1}x_{2}} &=&F_{x_{1}x_{2}x_{2}}\Psi +F_{x_{1}x_{1}x_{2}}\Psi
_{x_{1}},  \notag \\
\Psi _{x_{2}x_{2}} &=&F_{x_{2}x_{2}x_{2}}\Psi +F_{x_{1}x_{2}x_{2}}\Psi
_{x_{1}}.
\end{eqnarray}

Deformations of the structure constants generated by the DDA (47) in the
WDVV gauge are governed by the system (104) or (106) with the dropped
difference terms $\Delta _{j}A$ etc, i.e. by the system

\begin{eqnarray}
V_{11}+BU_{12}-U_{2}B_{2}-V_{2} =0,  \notag \\
U_{1}V_{12}-V_{1}U_{22}=0,  \notag \\
U_{11}U_{112}-U_{2}U_{12}-V_{11}B_{12}+BV_{12} =0.
\end{eqnarray}
Introducing W such that $B=W_{2}$, one rewrites this system as

\begin{eqnarray}
V_{11} =\left( UW_{2}+V-WU_{1}\right) _{2},\quad \left( \frac{V}{U}\right)
_{1}=\left( \frac{V_{1}}{U_{2}}\right) _{2},  \notag \\
\left( U_{1}U_{12}-V_{1}W_{22}\right) _{1} =\left( UU_{2}-WV_{1}\right) _{2}.
\end{eqnarray}

\section{ Deformations generated by three-dimensional Lie algebras and
discrete mappings}

\ Deformation moduli for the three-dimensional associative algebras
considered in sections 5 and 6 are two-dimensional. They are parametrized by
two discrete variables $x^{1}$ and $x^{2}$. One dimensional deformation
moduli arise naturally if one chooses a three-dimensional Lie algebra as
DDA. Among nine such nonequivalent Lie algebras only three generate discrete
deformations [24].

The first such DDA ($L_{2b}$) is defined by the commutation relations

\begin{equation}
\left[ p_{1},p_{2}\right] =0,\quad \left[ p_{1},x\right] =p_{1},\quad \left[
p_{2},x\right] =0.
\end{equation}
The relations (117) imply that

\begin{equation}
\left[ p_{1},\varphi (x)\right] =(T-1)\varphi (x)\cdot p_{1},\quad \left[
p_{2},\varphi (x)\right] =0
\end{equation}
where $T\varphi (x)=\varphi (x+1)$. Using this identity, one gets the CS

\begin{equation}
C_{1}TC_{2}=C_{2}C_{1}.
\end{equation}
For nondegenerate matrix $C_{1}$ one has

\begin{equation}
TC_{2}=C_{1}^{-1}C_{2}C_{1}.
\end{equation}

The CS (120) is the discrete version of the Lax equation and has similar
properties. It has three independent first integrals

\begin{equation}
I_{1}=trC_{2},\quad I_{2}=\frac{1}{2}tr(C_{2})^{2},\quad I_{3}=\frac{1}{3}%
tr(C_{2})^{3}
\end{equation}
and represents itself the compatibility condition for the linear problems

\begin{eqnarray}
\Phi C_{2} =\lambda \Phi ,  \notag \\
T\Phi =\Phi C_{1}.
\end{eqnarray}
Note that $\det C_{2}$ is the first integral too.

The CS (119) is the discrete dynamical system in the space of the structure
constants. For the two-dimensional algebra \textit{A} without the unite
element, i.e. when A=D=L=0 it \ has the form

\begin{eqnarray}
BTE+ETG &=&EB+MC,  \notag \\
BTM+ETN &=&E^{2}+MG,  \notag \\
CTE+GTG &=&BG+CN, \\
CTM+GTN &=&EG+NG  \notag
\end{eqnarray}%
where B and C are arbitrary functions. For nondegenerate matrix $C_{1},$
i.e. at $BG-CE\neq 0$ , in the resolved form it is

\begin{eqnarray}
TE &=&\frac{GM-EN}{BG-CE}C,\quad TG=B+\frac{BN-CM}{BG-CE}C,  \notag \\
TM &=&\frac{GM-EN}{BG-CE}G,\quad TN=E+\frac{BN-CM}{BG-CE}G.
\end{eqnarray}%
This system defines discrete deformations of the structure constants.
Formally, these deformations can be considered as the reduction $T_{2}$=
identity of the general case (90).

\ The second example is given the solvable DDA ($L_{4}$) defined by the
commutation relations

\begin{equation}
\left[ p_{1},p_{2}\right] =0,\quad \left[ p_{1},x\right] =p_{1},\quad \left[
p_{2},x\right] =p_{2}.
\end{equation}
For this DDA one has

\begin{equation}
\lbrack p_{j},\varphi (x)]=(T-1)\varphi (x)p_{j},\quad j=1,2
\end{equation}%
where $\varphi (x)$ is an arbitrary function and $T$ is the shift operator $%
T\varphi (x)=\varphi (x+1)$. With the use of (126) one arrives at the
following CS

\begin{equation}
C_{1}TC_{2}=C_{2}TC_{1}.
\end{equation}%
For nondegenerate matrix $C_{1}$ equation (127) is equivalent to the
equation $T(C_{2}C_{1}^{-1})=C_{1}^{-1}C_{2}$ or

\begin{equation}
TU=C_{1}^{-1}UC_{1}
\end{equation}
where $U\doteqdot C_{2}C_{1}^{-1}$. Using this form of the CS, one promptly
concludes that the CS (85) has three independent first integrals

\begin{equation}
I_{1}=tr(C_{2}C_{1}^{-1}),\quad I_{2}=\frac{1}{2}tr(C_{2}C_{1}^{-1})^{2},%
\quad I_{3}=\frac{1}{3}tr(C_{2}C_{1}^{-1})^{3}
\end{equation}
and is representable as the commutativity condition for the linear system

\begin{eqnarray}
\Phi C_{2}C_{1}^{-1} =\lambda \Phi ,  \notag \\
T\Phi =\Phi C_{1}.
\end{eqnarray}

For the two-dimensional algebra \textit{A} without unite element the CS (127
) is the system of four equations for six functions

\begin{eqnarray}
BTE+ETG =ETB+MTC,  \notag \\
BTM+ETN =ETE+MTG,  \notag \\
CTE+GTG =GTB+NTC,  \notag \\
CTM+GTN =GTE+NTG.
\end{eqnarray}

Chosing B and C as free functions and assuming that BG-CE$\neq 0$, one can
easily resolve (131) with respect to TE,TG,TM,TN. For instance, with B=C=1
one gets the following four-dimensional mapping

\begin{eqnarray}
TE =M-E\frac{M-N}{E-G},\quad TG=1+\frac{M-N}{E-G},  \notag \\
TM =N+(N-G)\frac{M-N}{E-G}-G\left( \frac{M-N}{E-G}\right) ^{2}, \\
TN =M+(1-E)\frac{M-N}{E-G}+\left( \frac{M-N}{E-G}\right) ^{2}.  \notag
\end{eqnarray}

\ In a similar manner one finds the CS associated with the DDA($\text{L}_{5}$%
) defined by the relations

\begin{equation}
\left[ p_{1},p_{2}\right] =0,\quad \left[ p_{1},x\right] =p_{1},\quad \left[
p_{2},x\right] =-p_{2}.
\end{equation}
Since in this case

\begin{equation}
\lbrack p_{1},\varphi (x)]=(T-1)\varphi (x)p_{1},\quad [p_{2},\varphi
(x)]=(T^{-1}-1)\varphi (x)p_{2}
\end{equation}
the CS takes the form

\begin{equation}
C_{1}TC_{2}=C_{2}T^{-1}C_{1}.
\end{equation}
For nondegenerate $C_{2}$ it is equivalent to

\begin{equation}
TV=C_{2}VC_{2}^{-1}
\end{equation}
where $V\doteqdot T^{-1}C_{1}\cdot C_{2}$. Similar to the previous case the
CS has three first integrals

\begin{equation}
I_{1}=tr(C_{1}TC_{2}),\quad I_{2}=\frac{1}{2}tr(C_{1}TC_{2})^{2},\quad I_{3}=%
\frac{1}{3}tr(C_{1}TC_{2})^{3}
\end{equation}
and is equivalent to the compatibility condition for the linear system

\begin{eqnarray}
(T^{-1}C_{1})C_{2}\Phi &=&\lambda \Phi ,  \notag \\
T\Phi &=&C_{2}\Phi .
\end{eqnarray}%
Note that the CS (135) is of the form (70) with $T_{1}=T,T_{2}=T^{-1}$.
Thus, the deformations generated by $\mathit{L}_{5}$ \ can be considered as
the reductions of the discrete deformations (70) under the constraint $%
T_{1}T_{2}C_{jk}^{n}=C_{jk}^{n}$.

\section{ Discrete oriented associativity equation}

\ For general discrete deformations , similar to the quantum deformations
[13], the global associativity condition $[C_{j},C_{k}]=0$ is not preserved
for all values of the deformation parameters. Deformations of associative
algebras for which the associativity condition is globally valid (
isoassociative deformations) form an important class of all possible
deformations [12-21]. Within the theories of Frobenius and F-manifolds
[12-14] and also for the coisotropic and quantum deformations [19,21] such
deformations are characterized by the existence of a set of functions $\Phi
^{l},l=1,...,N$ such that

\begin{equation}
C_{jk}^{l}=\frac{\partial ^{2}\Phi ^{l}}{\partial x^{j}\partial x^{k}},\quad
j,k,l=1,...,N.
\end{equation}

These functions obey the oriented associativity equation [87,13]

\begin{equation}
\sum_{m}\frac{\partial ^{2}\Phi ^{n}}{\partial x^{l}\partial x^{m}}\frac{%
\partial ^{2}\Phi ^{m}}{\partial x^{j}\partial x^{k}}-\sum_{m}\frac{\partial
^{2}\Phi ^{n}}{\partial x^{j}\partial x^{m}}\frac{\partial ^{2}\Phi ^{m}}{%
\partial x^{l}\partial x^{k}}=0,\quad j,k,l,n=1,...,N.
\end{equation}

Here we will present discrete versions of this equation. So, we consider the
N-dimensional associative algebra \textit{A,} deformations generated by the
DDA (48) and restrict ourselves to the isoassociative deformations for which

\begin{equation}
\lbrack C_{j}(x),C_{k}(x)]=0,\quad j,k=1,...,N.
\end{equation}
A class of solutions of the CS (70) equations is given by the formula

\begin{equation}
C_{j}=g^{-1}\Delta _{j}g
\end{equation}%
where g(x) is a matrix-valued function. Since $C_{jk}^{l}=C_{kj}^{l}$ one
has
\begin{equation}
\Delta _{j}g_{k}^{n}=\Delta _{k}g_{j}^{n},\qquad j,k,n=1,...,N
\end{equation}%
where $g_{k}^{n}$ are matrix elements of g. Hence

\begin{equation}
g_{k}^{n}=g_{0k}^{n}+\alpha \Delta _{k}\Phi ^{n},\qquad k,n=1,...,N
\end{equation}%
where $g_{0k}^{n}$ and $\alpha $ are arbitrary constants and $\Phi ^{n}$ are
functions. Substitution of (142) and (144) into (141) gives

\begin{equation}
\sum_{m,t}\Delta _{l}\Delta _{t}\Phi ^{n}\cdot (g^{-1})_{m}^{t}\Delta
_{j}\Delta _{k}\Phi ^{m}-\sum_{m,t}\Delta _{l}\Delta _{t}\Phi ^{n}\cdot
(g^{-1})_{m}^{t}\Delta _{j}\Delta _{k}\Phi ^{m}=0,\quad
\end{equation}%
Since in the continuous limit $\Delta _{j}\rightarrow \varepsilon \frac{%
\partial }{\partial x^{j}},g_{0k}^{n}=\delta _{k}^{n},\alpha =0,\varepsilon
\rightarrow 0$ the system (145) is reduced to (140), it represents a
discrete isoassociative version of the oriented associativity equation.

For the DDA (47) the CS (69) has a solution $C_{j}=g^{-1}T_{j}g$ and
symmetry of C implies that $g_{k}^{n}=T_{k}\Phi ^{n}$ where $\Phi ^{n}$ are
functions. Substitution of this expression into the associativity condition
(141) gives

\begin{equation}
\sum_{m,t}(T_{j}T_{m}\Phi ^{n})(g^{-1})_{t}^{m}T_{k}T_{l}\Phi
^{t}=\sum_{m,t}(T_{k}T_{m}\Phi ^{n})(g^{-1})_{t}^{m}T_{j}T_{l}\Phi ^{t}.
\end{equation}

Different discrete version of equation (140) arises if one relaxes the
condition (141) and requires that the following quasi-associativity condition

\begin{equation}
C_{l}T_{l}C_{j}=C_{j}T_{j}C_{l},\quad j,l=1,...,N
\end{equation}%
is valid for all values of deformation parameters . In this case the CS (70)
is reduced to the system

\begin{equation}
\Delta _{l}C_{j}-\Delta _{j}C_{l}=0,\quad j,l=1,...,N
\end{equation}
which implies the existence of the matrix-valued function $\Phi $ such that

\begin{equation}
C_{j}=\Delta _{j}\Phi ,\quad j=1,...,N.  \notag \\
\end{equation}
Since $C_{jk}^{l}=\Delta _{j}\Phi _{k}^{l}=C_{kj}^{l}$ one has

\begin{equation}
\Phi _{k}^{l}=\Delta _{k}\Phi ^{l},\quad l,k=1...,N  \notag \\
\end{equation}
where $\Phi ^{l},l=1,...,N$ are functions. So

\begin{equation}
C_{jk}^{l}=\Delta _{j}\Delta _{k}\Phi ^{l}.  \notag \\
\end{equation}

Finally, the quasi-associativity condition (147) takes the form

\begin{equation}
\sum_{m}\Delta _{j}\Delta _{k}T_{l}\Phi ^{m}\cdot \Delta _{l}\Delta _{m}\Phi
^{n}-\sum_{m}\Delta _{l}\Delta _{k}T_{j}\Phi ^{m}\cdot \Delta _{j}\Delta
_{m}\Phi ^{n}=0,\quad j,k,l,n=1,...,N.
\end{equation}
which is a discrete version of the oriented associativity equation (140).
Any solution of the systems (145), (146) and (149) defines discrete
deformation of the structure constants $C_{jk}^{l}.$

\section{Discrete deformations for a class of special associative algebras}

\ Several important discrete integrable equations arise as the CSs governing
discrete deformations of a class of associative algebras for which the
multiplication only of distinct elements of the basis is defined. For such
algebras the table of multiplication is of the form

\begin{equation}
\mathbf{P}_{j}\mathbf{P}_{k}=A_{jk}\mathbf{P}_{j}+B_{jk}\mathbf{P}_{k}+C_{jk}%
\mathbf{P}_{0},\quad j\neq k,\quad j,k=1,2,...,N.
\end{equation}

The commutativity of the basis implies that $A_{kj}=B_{jk},C_{jk}=C_{kj}$.
The algebra of the functions $f_{j}=\frac{a_{j}}{\lambda -\lambda _{j}}$
with simple poles in distinct points is an example of such algebra (see
(36)).

We choose the algebra (47) as DDA. Recall that the algebra of shifts $%
p_{j}=T_{j}$ \ is a realization of this DDA. Equations (62) and (45) in this
case take the form

\begin{equation}
\ p_{j}p_{k}\left| \Psi \right\rangle =A_{jk}p_{j}\left| \Psi \right\rangle
+B_{jk}p_{k}\left| \Psi \right\rangle +C_{jk}\left| \Psi \right\rangle
,\quad j\neq k
\end{equation}

and

\begin{equation}
\left( (p_{j}p_{k})p_{l}-p_{j}(p_{k}p_{l})\right) \left| \Psi \right\rangle
=\sum_{t}\Omega _{klj}^{t}p_{t}\left| \Psi \right\rangle ,\quad j\neq k\neq
l\neq j
\end{equation}
respectively. The corresponding CS is given by the following system of
equations

\begin{equation}
A_{kl}T_{l}B_{jk}=B_{jk}T_{j}A_{kl},
\end{equation}

\begin{equation}
B_{lj}T_{l}A_{jk}=A_{jk}T_{j}A_{kl}+A_{jl}T_{j}B_{kl}+T_{j}C_{kl},
\end{equation}

\begin{equation}
B_{jl}T_{j}B_{kl}=A_{lj}T_{l}A_{jk}+A_{lk}T_{l}B_{jk}+T_{l}C_{jk},
\end{equation}

\begin{equation}
C_{lj}T_{l}A_{jk}+C_{lk}T_{l}B_{jk}=C_{jk}T_{j}A_{kl}+C_{jl}T_{j}B_{kl}
\end{equation}
where all indices are distinct.

Denoting $\Phi _{j}\doteqdot T_{j}\left| \Psi \right\rangle ,\Phi
_{jk}\doteqdot T_{j}T_{k}\left| \Psi \right\rangle $, one rewrites (151) as
the system

\begin{equation}
\Phi _{jk}=A_{jk}\Phi _{j}+B_{jk}\Phi _{k}+C_{jk}\Phi ,\quad j\neq k.
\end{equation}
This system can be treated as the set of relations between the points $\Phi
,\Phi _{j},\Phi _{jk}$ connected by the shifts $\Phi _{j}=T_{j}\Phi ,\Phi
_{jk}=T_{j}T_{k}\Phi $.

The CS (153)- (156) and equations (151), (157) are invariant under the gauge
transformations (84) for which $\left| \Psi \right\rangle \rightarrow \left|
\widetilde{\Psi }\right\rangle =g^{-1}\left| \Psi \right\rangle ,\Phi
\rightarrow \widetilde{\Phi }=g^{-1}\Phi $ and

\begin{equation}
A_{jk}\rightarrow \widetilde{A}_{jk}=\frac{g_{j}}{g_{jk}}A_{jk},\quad
B_{jk}\rightarrow \widetilde{B}_{jk}=\frac{g_{k}}{g_{jk}}B_{jk},\quad
C_{jk}\rightarrow \widetilde{C}_{jk}=\frac{g}{g_{jk}}C_{jk}.
\end{equation}

Family of the structure constants connected by the transformations with
different g form orbits of gauge equivalent structure constants. These
orbits are characterized by the $N(N-1)$ gauge invariants

\begin{equation}
I_{jk}^{1}=\frac{B_{jk}\cdot T_{j}^{-1}A_{jk}}{A_{jk}\cdot T_{k}^{-1}B_{jk}}%
,\quad I_{jk}^{2}=\frac{C_{jk}}{B_{jk}\cdot T_{j}^{-1}A_{jk}},\quad j\succ
k,j,k=1,...,N
\end{equation}

or

\begin{equation}
\widetilde{I}_{jk}^{1}=\frac{B_{jk}\cdot T_{j}^{-1}A_{jk}}{C_{jk}},\quad
\widetilde{I}_{jk}^{2}=\frac{A_{jk}\cdot T_{k}^{-1}B_{jk}}{C_{jk}},\quad
j\succ k.
\end{equation}

These invariants are the discrete version of the well-known gauge invariants
for the standard Laplace equations (see e.g. [88]). The invariants (160)
coincide with those introduced earlier in [89] up to the trivial
redefinitions.

\ Similar to the continuous case (see [88]) these invariants have also the
meaning of the defects of the factorizability of the discrete operators,
namely,

\begin{eqnarray}
T_{j}T_{k}-A_{jk}T_{j}-B_{jk}T_{k}-C_{jk}
=(T_{j}-B_{jk})(T_{k}-(T_{j}^{-1}A_{jk}))+C_{jk}(\widetilde{I}_{jk}^{1}-1)=
\notag \\
=(T_{j}-A_{jk})(T_{k}-(T_{k}^{-1}B_{jk}))+C_{jk}(\widetilde{I}_{jk}^{2}-1).
\end{eqnarray}

The orbits of gauge transformations (158) have one distinguished element.
Indeed, it is easy to see that choosing $g=$  $\widehat{\Phi }$ where  $%
\widehat{\Phi }$ is  a solution of the system (157), one gets the structure
constants $\widetilde{A}_{jk},$ $\widetilde{B}_{jk},\widetilde{C}_{jk}$
which obey the relation

\begin{equation}
\widetilde{A}_{jk}+\widetilde{B}_{jk}+\widetilde{C}_{jk}=1,\quad j\neq k.
\end{equation}
In this gauge the relations (157) and corresponding configuration of the
points are invariant under translations in space.

\ The multidimensional multiplication table (150) and the CS (153-156)
are,in fact, the systems of three-dimensional irreducible subsystems glued
all together. Indeed, for any three distinct indices j,k,l the corresponding
subtable (150) and subsystem of (153-156) are closed with other deformation
variables playing the role of parameters. So, the study of the CS (153-156)
is reduced basically to the study of the three-dimensional case. Choosing
any three distinct indices j,k,l and denoting them 1,2,3, we present a
corresponding subtable of multiplication as

\begin{equation}
\mathbf{P}_{1}\mathbf{P}_{2}=A\mathbf{P}_{1}+B\mathbf{P}_{2}+L\mathbf{P}_{0},
\end{equation}%
\begin{equation}
\mathbf{P}_{1}\mathbf{P}_{3}=C\mathbf{P}_{1}+D\mathbf{P}_{3}+M\mathbf{P}_{0},
\end{equation}%
\begin{equation}
\mathbf{P}_{2}\mathbf{P}_{3}=E\mathbf{P}_{2}+G\mathbf{P}_{3}+N\mathbf{P}_{0}.
\end{equation}

The subsystem of the CS for the structure constants A,B,...,N is given by
the system of equations

\begin{eqnarray}
\frac{A_{3}}{A} &=&\frac{C_{2}}{C},\qquad \frac{B_{3}}{B}=\frac{E_{1}}{E}%
,\qquad \frac{D_{2}}{D}=\frac{G_{1}}{G}, \\
(A_{3}-E_{1})L+B_{3}N-G_{1}M &=&0, (A_{3}-G_{1})D-E_{1}A-N_{1}=0, \\
(A_{3}-G_{1})D+B_{3}G+L_{3} &=&0, (C_{2}-E_{1})L+D_{2}N-G_{1}M=0, \\
(C_{2}-E_{1})B+D_{2}E+M_{2} &=&0, C_{2}L-A_{3}M+(D_{2}-B_{3})N=0
\end{eqnarray}%
where we denote $A_{j}=T_{j}A,B_{j}=T_{j}B$ and so on. \ \ For the
distinguished gauge (162) one has

\begin{equation}
A+B+L=1,
\end{equation}%
\begin{equation}
C+D+M=1,
\end{equation}%
\begin{equation}
E+G+N=1.
\end{equation}

One may note the coincidence of these relations with those which define the
so-called barycentric coordinates of a point in the plane of a given
triangle (see e.g. [90,91]).

\section{ Menelaus relation as associativity condition and Menelaus
deformations.}

\ Deformations of the associative algebras of the type (150) but without
unite element are of particular interest. In this case the table of
multiplication is given by (15) with $C_{jk}=0.$ As in the previous section
we choose the algebra (47) as DDA.

For the three-dimensional irreducible subalgebra the multiplication table is
given by (163-165) with L=M=N=0 while equations (153) are of the form

\begin{equation}
\Phi _{12}=A\Phi _{1}+B\Phi _{2},\quad \Phi _{13}=C\Phi _{1}+D\Phi
_{3},\quad \Phi _{23}=E\Phi _{2}+G\Phi _{3}.
\end{equation}
The distinguished gauge is then defined by the relations

\begin{equation}
A+B=1,\quad C+D=1,\quad E+G=1.
\end{equation}
The associativity conditions

\begin{equation}
\mathbf{P}_{1}(\mathbf{P}_{2}\mathbf{P}_{3})=\mathbf{P}_{2}(\mathbf{P}_{3}%
\mathbf{P}_{1})=\mathbf{P}_{3}(\mathbf{P}_{1}\mathbf{P}_{2})
\end{equation}
in this case are equivalent to the system

\begin{equation}
(A-G)C-EA=0,\quad (A-G)D+BG=0,\quad (C-E)B+DE=0.
\end{equation}

This system is a rather special one. First, equations (176) imply that
AED+BCG=0. Then it is easy to check that one has the following

\textbf{Proposition 3.} For nonzero A, B,...,G the associativity conditions
(176) are equivalent to the equation

\begin{equation}
AED+BCG=0
\end{equation}
and one of equations (176), for instance, the equation

\begin{equation}
(A-G)C-EA=0.\Box
\end{equation}

This form of associativity conditions is of interest for several reasons.
One of them is connected with gauge transformations. It was noted in the
previous section that the associativity conditions and, in particular,
conditions (176) are not invariant under general gauge transformations. The
form (177), (178) of associativity conditions has a special property.
Namely, due to the relation

\begin{equation}
\widetilde{A}\widetilde{E}\widetilde{D}+\widetilde{B}\widetilde{C}\widetilde{%
G}=\frac{g_{1}g_{2}g_{3}}{g_{12}g_{23}g_{13}}(AED+BCG)
\end{equation}%
the condition (177) is invariant under gauge transformations. So, this
relation is a characteristic one for the orbits of gauge equivalent
structure constants.

Moreover, one can show that the condition (177) guarantees that the set of
constants A,B,...,G can be converted into the set of constants $\widetilde{A}%
,\widetilde{B},...,\widetilde{G}$ \ obeying the associativity conditions
(176) by the gauge transformation with $g=\widehat{\Phi }$ where $\widehat{%
\Phi }$ is a solution of equations (173).

Then, of one treats equations (173) as the relations between six points $%
\Phi _{1},\Phi _{2},\Phi _{3},$ $\Phi _{12},\Phi _{23},\Phi _{13}$ on the
(complex) plane then the l.h.s. of (177) coincides with the determinant of
the matrix of transformation from the set of three points $\Phi _{1},\Phi
_{2},\Phi _{3}$ to the set $\Phi _{12},\Phi _{23},\Phi _{13}$. So, for
associative algebras such transformation is singular.

\ The relation between six points $\Phi _{1},\Phi _{2},\Phi _{3},$ $\Phi
_{12},\Phi _{23},\Phi _{13}$ defined by (173) has really a remarkable
geometrical meaning in the distinguished gauge (176). First, the relations
(173), in virtue of the conditions (174), mean that three points $\Phi
_{1},\Phi _{2},\Phi _{12}$ are collinear as well as the sets of points $\Phi
_{1},\Phi _{3},\Phi _{13}$ and $\Phi _{2},\Phi _{3},\Phi _{23}$. \ Then the
relations (177), (178) imply that the points $\Phi _{12},\Phi _{13},\Phi
_{23}$ are collinear too, i.e.
\begin{equation}
\Phi _{12}=\frac{A}{C}\Phi _{13}+\frac{B}{E}\Phi _{23}
\end{equation}%
with $\frac{A}{C}+\frac{B}{E}=1$. \ Thus, in the gauge (174) the relations
(173) describe the set of four triples $(\Phi _{1},\Phi _{2},\Phi
_{12}),(\Phi _{1},\Phi _{3},\Phi _{13}),(\Phi _{2},\Phi _{3},\Phi _{23})$
and $(\Phi _{12},\Phi _{13},\Phi _{23})$ of collinear points. It is nothing
but the celebrated Menelaus configuration of the classical geometry (figure
1) ( see e.g. [90], [91]). Due to (174) the relations (173) and Menelaus
configuration are translationally invariant.
\begin{figure}[t]
\begin{center}
\includegraphics[width=8cm, angle=0]{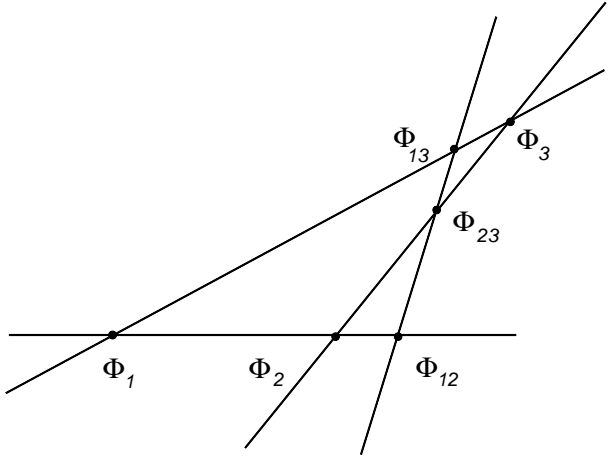}
\end{center}
\caption{{\protect\small Menelaus configuration.}}
\label{Figure1}
\end{figure}

The relations (173) and (174) allow us to express A,B,...,G in terms of $%
\Phi $. One gets

\begin{eqnarray}
A &=&\frac{\Phi _{12}^{M}-\Phi _{2}^{M}}{\Phi _{1}^{M}-\Phi _{2}^{M}},\quad
B=- \frac{\Phi _{12}^{M}-\Phi _{1}^{M}}{\Phi _{1}^{M}-\Phi _{2}^{M}},\quad C=%
\frac{\Phi _{13}^{M}-\Phi _{3}^{M}}{\Phi _{1}^{M}-\Phi _{3}^{M}},  \notag \\
D &=&-\frac{\Phi _{13}^{M}-\Phi _{1}^{M}}{\Phi _{1}^{M}-\Phi _{3}^{M}},\quad
E=\frac{\Phi _{23}^{M}-\Phi _{3}^{M}}{\Phi _{2}^{M}-\Phi _{3}^{M}},\quad G=-%
\frac{\Phi _{23}^{M}-\Phi _{2}^{M}}{\Phi _{2}^{M}-\Phi _{3}^{M}}
\end{eqnarray}
where we denote by $\Phi ^{M}$ solution of the system (173), (174). In such
a parametrization of A, B,...,G the relations (177), (178) are equivalent to
the single equation

\begin{equation}
\frac{(\Phi _{1}^{M}-\Phi _{12}^{M})(\Phi _{2}^{M}-\Phi _{23}^{M})(\Phi
_{3}^{M}-\Phi _{13}^{M})}{(\Phi _{12}^{M}-\Phi _{2}^{M})(\Phi _{23}^{M}-\Phi
_{3}^{M})(\Phi _{13}^{M}-\Phi _{1}^{M})}=-1.
\end{equation}

It is the famous Menelaus relation (see [90], [91]) which guarantees that
for any three points $\Phi _{1},\Phi _{2},\Phi _{3}$ on the plane the points
$\Phi _{12},\Phi _{13},\Phi _{23}$ are collinear. In our formulation the
Menelaus relation (182) is nothing else than the associativity conditions
(177), (178) written in terms of $\Phi ^{M}$. Thus, the Menelaus theorem is
intimately connected with the associative algebra (163), (165) with L=M=N=0
in the distinguished gauge (174). Amazingly, this interpretation seems to be
not that distant from the old known algebraic proofs of the Menelaus theorem
where the relation of the type (177) already has appeared [90].

Comparing equations (182) and (26), one concludes that the Menelaus relation
and NSF for the Schwarzian KP hierarchy coincide for given six points $\Phi
_{1},\Phi _{2},\Phi _{3},$ $\Phi _{12},\Phi _{23},\Phi _{13}$ [48]. In the
paper [48] this coincidence has been extended to the discrete equation thus
defining the Menelaus lattice.

\ In our approach discrete deformations of the algebra (163)-(166) and,
hence, of the Menelaus configuration is governed by the CS

\begin{equation}
\frac{A_{3}}{A} =\frac{C_{2}}{C},\qquad \frac{B_{3}}{B}=\frac{E_{1}}{E}%
,\qquad \frac{D_{2}}{D}=\frac{G_{1}}{G},
\end{equation}
\begin{equation}
(A_{3}-G_{1})C-E_{1}A =0,
\end{equation}
\begin{equation}
(A_{3}-G_{1})D+B_{3}G =0,
\end{equation}
\begin{equation}
(C_{2}-E_{1})B+D_{2}E =0.
\end{equation}
Equations (184), (185) imply that

\begin{equation}
AE_{1}D+B_{3}CG=0  \notag \\
\end{equation}
which is equivalent to the relation (177) in virtue of the second equation
(183). Thus, the Menelaus relation (182) is preserved by deformations.
Discrete deformations of the Menelaus configuration given by equations
(183)-(186) generate a lattice on the plane . It is a straighforward check
that equations (183)-(186) rewritten in terms of the ''shape'' parameters $%
\alpha =-\frac{B}{A},\beta =-\frac{G}{E},\gamma =-\frac{C}{D}$ coincide with
those derived in [48] (Theorem 3). Thus, the Menelaus lattice represents a
realization of discrete deformations of the associative algebra (163)-(165)
with L=M=N=0 in the gauge (174).

In [48] it was shown that under the constraint $\Phi _{23}=\Phi $ the
Menelaus relation (182) is reduced to the Schwarzian discrete KdV equation
(9). So, the under this constraint one has a subclass of deformations
governed by the discrete KdV equation (9).

\section{KP configurations, discrete KP deformations and their gauge
equivalence to Menelaus configurations}

\ There is an another distinguished gauge on the orbits under consideration.
It is given by the relations

\begin{equation}
A+B=0,\quad C+D=0,\quad E+G=0
\end{equation}
for which equations (173) take the form

\begin{equation}
\Phi _{12}^{KP}=A(\Phi _{1}^{KP}-\Phi _{2}^{KP}),\quad \Phi
_{13}^{KP}=C(\Phi _{1}^{KP}-\Phi _{3}^{KP}),\quad \Phi _{23}^{KP}=E(\Phi
_{2}^{KP}-\Phi _{3}^{KP}).
\end{equation}
We shall refer to this gauge as the KP gauge for the reason which will be
clarified below. In this gauge the relation (177) becomes a trivial identity
and , hence, the associativity conditions are reduced to the single equation

\begin{equation}
AC+EC-AE=0.
\end{equation}

\textbf{Proposition 4.} In the KP gauge (187) the CS (183)- (186) is
equivalent to the associativity condition (189) and equations

\begin{equation}
\frac{A_{3}}{A}=\frac{C_{2}}{C}=\frac{E_{1}}{E}.
\end{equation}

Proof. In the gauge (187) equations (183) are reduced to equations (190)
while equations (184)- (186) are equivalent to the single equation $%
A_{3}C+E_{1}C-AE_{1}=0$. Due to (190) this equation is equivalent to the
associativity condition(189).$\Box $

Equations (190) imply the existence of a function $\tau $ such that

\begin{equation}
A=-\frac{\tau _{1}\tau _{2}}{\tau \tau _{12}},\quad C=-\frac{\tau _{1}\tau
_{3}}{\tau \tau _{13}},\quad E=-\frac{\tau _{2}\tau _{3}}{\tau \tau _{23}}.
\end{equation}
Substitution of these expressions into (189) gives the Hirota bilinear
equation for the KP hierarchy, i.e.

\begin{equation}
\tau _{1}\tau _{23}-\tau _{2}\tau _{13}+\tau _{3}\tau _{12}=0.
\end{equation}

This fact justifies the name of the gauge (187). We would like to emphasize
that the Hirota-Miwa equation (192) is nothing but the associativity
condition (189) with the structure constants A, C, E parametrized by $\tau $%
-function.

\ Equations (188) with A, C, E of the form (191) coincide with well-known
linear problems for the Hirota-Miwa bilinear equation [92,35]. The
parametrization (191) suggests to rewrite equation (188) in the gauge
equivalent form

\begin{equation}
\widehat{\Phi }_{12}^{KP}=\frac{\tau _{1}}{\tau }\widehat{\Phi }_{2}^{KP}-%
\frac{\tau _{2}}{\tau }\widehat{\Phi }_{1}^{KP},\quad \widehat{\Phi }%
_{13}^{KP}=\frac{\tau _{1}}{\tau }\widehat{\Phi }_{3}^{KP}-\frac{\tau _{3}}{%
\tau }\widehat{\Phi }_{1}^{KP},\quad \widehat{\Phi }_{23}^{KP}=\frac{\tau
_{2}}{\tau }\widehat{\Phi }_{3}^{KP}-\frac{\tau _{3}}{\tau }\widehat{\Phi }%
_{2}^{KP}
\end{equation}
where $\widehat{\Phi }^{KP}=\tau \Phi ^{KP}$. \ The condition of
compatibility for the system (193) is equivalent to equation (192) too.
Equations (193) imply also that

\begin{equation}
\widehat{\Phi }_{1}^{KP}\widehat{\Phi }_{23}^{KP}-\widehat{\Phi }_{2}^{KP}%
\widehat{\Phi }_{13}^{KP}+\widehat{\Phi }_{3}^{KP}\widehat{\Phi }_{12}^{KP}-0
\end{equation}
that coincides with equation (192). So, one can choose $\widehat{\Phi }^{KP}=%
\widehat{\tau }$ where $\widehat{\tau }$ is a solution of the Hirota-Miwa
equation (192). Thus, one has

\begin{equation}
\Phi ^{KP}=\frac{\widehat{\tau }}{\tau }
\end{equation}
which is again \ the well-known formula from the theory of the KP hierarchy.

Geometrical configuration on the plane formed by six points $\Phi
_{1}^{KP},\Phi _{2}^{KP},\Phi _{3}^{KP},$ $\Phi _{12}^{KP},\Phi
_{23}^{KP},\Phi _{13}^{KP}$ with real A, C, E is of interest too. We, first,
observe that the points $\Phi _{12}^{KP},\Phi _{23}^{KP},\Phi _{13}^{KP}$
lie on the straight lines passing through the origin 0 and parallel to the
straight lines passing through the points ($\Phi _{1}^{KP},\Phi
_{2}^{KP}),(\Phi _{1}^{KP},\Phi _{3}^{KP}),$ $(\Phi _{2}^{KP},\Phi _{3}^{KP})
$, respectively. Then, due to the associativity condition (189) the points $%
\Phi _{12}^{KP},\Phi _{23}^{KP},\Phi _{13}^{KP}$ are collinear. Indeed,
equations (188) imply that

\begin{equation}
\frac{1}{C}\Phi _{13}^{KP}-\frac{1}{A}\Phi _{12}^{KP}-\frac{1}{E}\Phi
_{23}^{KP}=0
\end{equation}
while the relation (189) is equivalent to the condition $\frac{1}{C}-\frac{1%
}{A}-\frac{1}{E}=0.$Thus, the points $\Phi _{1}^{KP},\Phi _{2}^{KP},\Phi
_{3}^{KP},$ $\Phi _{12}^{KP},\Phi _{23}^{KP},\Phi _{13}^{KP}$ form the KP
configuration on the complex plane shown on the figure 2.

\begin{figure}[tbp]
\begin{center}
\includegraphics[width=12cm, angle=0]{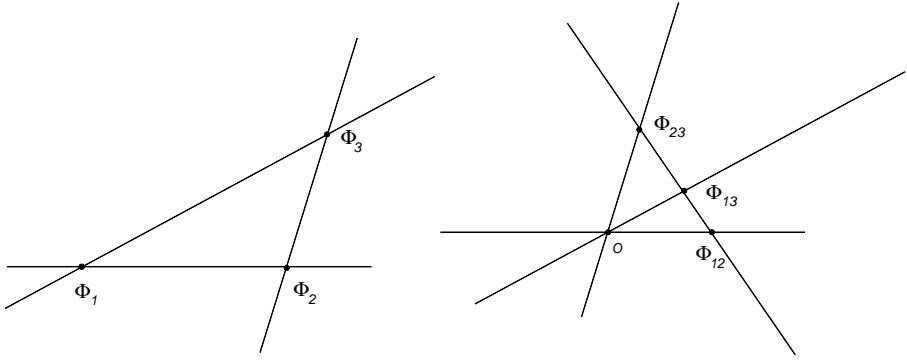}
\end{center}
\caption{{\protect\small $KP$ configuration.}}
\label{Figure2}
\end{figure}

\ The associativity condition (189) provides us also with the relation
between the directed lengths for the KP configuration. Indeed, expressing A,
C, E from (188) in terms of $\Phi ^{KP}$ and substituting into (189), one
gets

\begin{equation}
\frac{\Phi _{1}^{KP}-\Phi _{2}^{KP}}{\Phi _{12}^{KP}}+\frac{\Phi
_{2}^{KP}-\Phi _{3}^{KP}}{\Phi _{23}^{KP}}+\frac{\Phi _{3}^{KP}-\Phi
_{1}^{KP}}{\Phi _{31}^{KP}}=0.
\end{equation}
Since for real A, C, E $\frac{\Phi _{1}^{KP}-\Phi _{2}^{KP}}{\Phi _{12}^{KP}}%
=\frac{\left| \Phi _{1}^{KP}-\Phi _{2}^{KP}\right| }{\left| \Phi
_{12}^{KP}\right| }$ etc the formula (197) represents the relation between
the directed lengths $\left| \Phi _{1}^{KP}-\Phi _{2}^{KP}\right| $ of the
interval $(\Phi _{1},\Phi _{2})$ etc. In contrast to the Menelaus case the
relations (188), (197) and KP configuration are not invariant with respect
to the displacements on the plane.

The Menelaus and KP configurations look quite different. For instance, the
points $\Phi _{1},\Phi _{2},\Phi _{12}$ etc are collinear in the Menelaus
case and they are not in the KP case. Nevertheless, they are closely
connected, namely, they are gauge equivalent to each other. To demonstrate
this let us consider the gauge transformation $\Phi ^{KP}\rightarrow
\widetilde{\Phi }^{KP}=g^{-1}\Phi ^{KP}$. Under this transformation

\begin{eqnarray}
\widetilde{A}=\frac{g_{1}}{g_{12}}A^{KP},\quad \widetilde{B}=-\frac{g_{2}}{%
g_{12}}A^{KP},\quad \widetilde{C}=\frac{g_{1}}{g_{13}}C^{KP},  \notag \\
\widetilde{D}=-\frac{g_{3}}{g_{13}}C^{KP},\quad \widetilde{E}=\frac{g_{2}}{%
g_{23}}E^{KP},\quad \widetilde{G}=-\frac{g_{3}}{g_{23}}E^{KP}
\end{eqnarray}

and

\begin{equation}
\widetilde{A}+\widetilde{B}=\frac{g_{1}-g_{2}}{g_{12}}A^{KP},\quad
\widetilde{C}+\widetilde{D}=\frac{g_{1}-g_{3}}{g_{13}}C^{KP},\quad
\widetilde{E}+\widetilde{G}=\frac{g_{2}-g_{3}}{g_{23}}E^{KP}.  \notag \\
\end{equation}

Choosing $g=\widehat{\Phi }^{KP}$ where $\widehat{\Phi }^{KP}$ is a solution
of equations (188) with the same $A^{KP},C^{KP},E^{KP}$, one gets

\begin{equation}
\widetilde{A}+\widetilde{B}=1,\quad \widetilde{C}+\widetilde{D}=1,\quad
\widetilde{E}+\widetilde{G}=1.
\end{equation}

Thus, $\widetilde{\Phi }^{KP}=\frac{\Phi ^{KP}}{\widehat{\Phi }^{KP}}=\Phi
^{M}$ and so the KP configuration is converted into the Menelaus
configuration. In geometric terms this gauge transformation is the local (
depending on the point) homothetic transformation. So, in order to construct
the Menelaus configuration out of the KP we need two KP configurations with
the same A, C, E. Similarly, the formula (195) shows us that to construct KP
configuration one needs two configurations of six points defined by the
Hirota-Miwa equation (192).

Another way to demonstrate the gauge equivalence between Menelaus and KP
configurations is based on the formula (27). Eliminating $\Psi ^{\ast }$,
one can rewrite this equation as [65]

\begin{equation}
\frac{\widetilde{\Delta }_{j}\Phi }{\widetilde{\Delta }_{k}\Phi }=\frac{%
T_{j}^{-1}\Psi }{T_{k}^{-1}\Psi },\quad j\neq k,j,k=1,2,3
\end{equation}
where $\widetilde{\Delta }_{j}=T_{j}^{-1}-1.$ These equations imply the
following

\begin{equation}
\Phi _{jk}=-\frac{\Psi _{k}}{\Psi _{j}-\Psi _{k}}\Phi _{j}+\frac{\Psi _{j}}{%
\Psi _{j}-\Psi _{k}}\Phi _{k},\quad j\neq k,j,k=1,2,3.
\end{equation}
Thus, $\Phi =\Phi ^{M}$ and the formulae (201) give us the parametrization
of A, B, ...,G in terms of $\Psi $ in the Menelaus gauge.

Performing the gauge transformation $\Phi =\Psi \widetilde{\Phi }$, one gets
the equations

\begin{equation}
\widetilde{\Phi }_{jk}=-\frac{\Psi _{j}\Psi _{k}}{\Psi _{jk}(\Psi _{j}-\Psi
_{k})}\widetilde{\Phi }_{j}+\frac{\Psi _{j}\Psi _{k}}{\Psi _{jk}(\Psi
_{j}-\Psi _{k})}\widetilde{\Phi }_{k},\quad j\neq k,j,k=1,2,3.
\end{equation}
So, $\widetilde{\Phi }=\Phi ^{KP}$ and the formulae (202) provide us with
the parametrization of A, B,..., G in the KP case in terms of $\Psi $.

It is an easy check that the ratio $\frac{\widetilde{\Phi }}{\widehat{\Phi }}
$ of two solutions of the system (202) obeys the Menelaus system (201) with $%
\Psi =\frac{1}{\widehat{\Phi }}$ in agreement with the formula $\Phi ^{M}=%
\frac{\Phi ^{KP}}{\widehat{\Phi }^{KP}}$. Then, using parametrization of $%
A_{jk}$ and $B_{jk}$ given by the formulae (201) and (202), it is not
difficult to show that the invariants $I_{jk}^{1}$ (159) are equal in
Menelaus and KP gauges.

Note also that in the parametrization (201) of the Menelaus A, B,...,G the
associativity conditions (177), (178) are satisfied identically. In the KP
case (202) these associativity conditions are satisfied in virtue of
equation (197) for $\Phi ^{KP}=\frac{1}{\Psi }$. The collinearity conditions
for points $\Phi _{12},\Phi _{13},\Phi _{23}$ are obviously satisfied in
both cases in parametrizations (201) and (202).

At last, the Menelaus figure 1 and KP figure 2 are converted to each other
by the gauge transformation ( local homothety) $\Phi ^{M}=\Psi \Phi ^{KP}.$

\section{ Multidimensional Menelaus and KP configurations and deformations}

Now let us consider the N-dimensional algebra (150) with $C_{jk}=0$ in the
Menelaus gauge $A_{jk}+B_{jk}=1,j\neq k.$ The associativity conditions in
this case take the form

\begin{equation}
A_{jk}A_{jl}=A_{kl}A_{jk}+A_{jl}B_{kl},\quad
B_{kl}B_{jl}=A_{jk}A_{lj}+B_{jk}A_{lk}
\end{equation}%
where all indices are distinct. Multiplying the first of these equations by $%
B_{jl}$ , second by $A_{jl}$ and subtracting, one gets

\begin{equation}
A_{jk}A_{kl}B_{jl}+A_{jl}B_{jk}B_{kl}=0.
\end{equation}%
with all distinct indices.

For N=3 it is just the Menelaus relation (177). For arbitrary N the above
Menelaus type relations imply

\begin{equation}
\frac{B_{12}}{A_{12}}\frac{B_{23}}{A_{23}}\frac{B_{34}}{A_{34}}...\frac{%
A_{1N}}{B_{1N}}=(-1)^{N}.
\end{equation}

From equations (157) with $C_{jk}=0$ in the Menelaus gauge one gets

\begin{equation}
A_{jk}=\frac{\Phi _{jk}-\Phi _{k}}{\Phi _{j}-\Phi _{k}},\quad B_{jk}=-\frac{%
\Phi _{jk}-\Phi _{j}}{\Phi _{j}-\Phi _{k}},\quad j,k=1,...,N.
\end{equation}
Substituting these expressions into (205), one obtains the relation

\begin{equation}
\frac{(\Phi _{1}-\Phi _{12})(\Phi _{2}-\Phi _{23})\cdot \cdot \cdot (\Phi
_{N}-\Phi _{1N})}{(\Phi _{12}-\Phi _{2})(\Phi _{23}-\Phi _{3})\cdot \cdot
\cdot (\Phi _{1N}-\Phi _{1})}=(-1)^{N}.
\end{equation}

It is the generalization of the Menelaus relation to N-gons on the plane
where $\Phi _{1},...,\Phi _{N}$ are vertices of the N-gon and $\Phi
_{12},\Phi _{23},...,\Phi _{N1}$ are points of intersections of a straight
line with the corresponding sides of N-gon [93] (see also [50]). Again it is
just the associativity condition (204).

Deformations of the N-dimensional algebra (150) and N-gon Menelaus
configuration are governed by the CS

\begin{eqnarray}
A_{kl}T_{l}B_{jk} &=&B_{jk}T_{j}A_{kl},  \notag \\
B_{lj}T_{l}A_{jk} &=&A_{jk}T_{j}A_{kl}+A_{jl}T_{j}B_{kl},  \notag \\
B_{jl}T_{j}B_{kl} &=&A_{lj}T_{l}A_{jk}+A_{lk}T_{l}B_{jk}
\end{eqnarray}
with all distinct indices j,k,l.

\ In the KP gauge $A_{jk}+B_{jk}=0$ the associativity conditions for the
N-dimensional algebra (150) with $C_{jk}=0$ are reduced to the system

\begin{equation}
A_{jk}A_{jl}-A_{kl}A_{jk}+A_{jl}A_{kl}=0
\end{equation}
with distinct indices j,k,l. Associated geometrical configurations on the
plane are of interest too. For example, at N=4 one has four connected KP
configuations of the type shown in the figure 2. For the triples of points $%
(\Phi _{1,}\Phi _{2},\Phi _{3}),(\Phi _{1,}\Phi _{2},\Phi _{4}),$ $(\Phi
_{1,}\Phi _{3},\Phi _{4}),(\Phi _{2,}\Phi _{3},\Phi _{4})$ one has four
triples
\begin{equation}
(\Phi _{12,}\Phi _{13},\Phi _{23}),(\Phi _{12,}\Phi _{14},\Phi _{24}),(\Phi
_{13,}\Phi _{14},\Phi _{34}),(\Phi _{23,}\Phi _{24},\Phi _{34})  \notag \\
\end{equation}
of collinear points. The latter form the classical Menelaus configuration (
figure 1).

Deformations of such multi-KP configurations are governed by the CS

\begin{equation}
A_{kl}T_{l}A_{jk} =A_{jk}T_{j}A_{kl,}
\end{equation}
\begin{equation}
A_{jk}T_{j}A_{kl}-A_{jl}T_{j}A_{kl}-A_{jl}T_{l}A_{jk} =0.
\end{equation}

Similar to the case N=3 this system is equivalent to the associativity
condition (209) and equation (210). Equations (210) imply the existence of
the function $\tau $ such that

\begin{equation}
A_{jk}=-\frac{\tau _{j}\tau _{k}}{\tau \tau _{jk}}.
\end{equation}
Substitution of this expression into (209) gives

\begin{equation}
\tau _{j}\tau _{kl}-\tau _{k}\tau _{jl}+\tau _{l}\tau _{jk}=0,\quad
j,k,l=1,...,N
\end{equation}
with the distinct indices j,k,l. This is the multi-dimensional version of
the Hirota-Miwa equation (192).

One has also the multi-dimensional versions of the relations (197),
parametrizations of $A_{jk}$ and $B_{jk}$ given by (201) and (202) as well
as the gauge equivalence between the multi-Menelaus and multi-KP
configurations.

\section{Discrete Darboux system and discrete BKP Hirota-Miwa equation.}

\ Now we will turn back to the algebra (150) with unite element. The table
of multiplication for the three-dimensional irreducible subalgebra is given
by (163)- (165) and the distinguished gauge is defined by relations (170)-
(172). Geometrically the triples of structure constants (A, B, L), (C, D,
M), (E, G, N) are the barycentric ( or normalized ) cooordinates of the
points $\Phi _{12},\Phi _{13},\Phi _{23}$ in the plane of the given
triangles with the vertices in points $(\Phi _{1},\Phi _{2},\Phi ),(\Phi
_{1},\Phi _{3},\Phi ),$ $(\Phi _{2},\Phi _{3},\Phi )$, respectively.

Equations (166) from the CS imply that there exist three functions U, V, W
such that

\begin{equation}
A=\frac{U_{2}}{U},\quad B=\frac{V_{1}}{V},\quad C=\frac{U_{3}}{U},\quad D=%
\frac{W_{1}}{W},\quad E=\frac{V_{3}}{V},\quad G=\frac{W_{2}}{W}.
\end{equation}
In terms of the functions $H^{1},H^{2},H^{3}$ defined by

\begin{equation}
U=H_{1}^{1},\quad V=H_{2}^{2},\quad W=H_{3}^{3}
\end{equation}
the CS (161)-(164) under the constraints (170)-(172) takes the form

\begin{equation}
H_{lk}^{j}-\frac{H_{kl}^{k}}{H_{k}^{k}}H_{k}^{j}-\frac{H_{kl}^{l}}{H_{l}^{l}}%
H_{l}^{j}+\frac{H_{lk}^{k}}{H_{k}^{k}}H^{j}+\frac{H_{lk}^{l}}{H_{l}^{l}}%
H^{j}-H^{j}=0
\end{equation}
or equivalently

\begin{equation}
\Delta _{l}\Delta _{k}H^{j}-\frac{\Delta _{l}H_{k}^{k}}{H_{k}^{k}}\cdot
\Delta _{k}H^{j}-\frac{\Delta _{k}H_{l}^{l}}{H_{l}^{l}}\cdot \Delta
_{l}H^{j}=0
\end{equation}%
where indices j,k,l=1,2,3 are all distinct. It is the well-known discrete
Darboux system which was first derived in [74]. It describes the
deformations of barycentric coordinates discussed above.

For the general N-dimensional case (150) with the constraints

\begin{equation}
A_{jk}+B_{jk}+C_{jk}=1,\quad j\neq k,
\end{equation}%
one has

\begin{equation}
A_{jk}=\frac{H_{kj}^{k}}{H_{k}^{k}},\qquad B_{jk}=\frac{H_{kj}^{j}}{H_{j}^{j}%
},\qquad C_{jk}=1-\frac{H_{kj}^{k}}{H_{k}^{k}}-\frac{H_{kj}^{j}}{H_{j}^{j}}
\end{equation}%
and the CS is given by N-dimensional systems (216) or (217).

So, the discrete Darboux system governs discrete deformations of the
structure constants for the distinguished elements of the gauge equivalency
orbits.

\ Discrete Darboux system (217) is an important system in discrete geometry,
in particular, in the theory of quadrilateral lattices (see e.g. [94],
[95]). The above result shows that the theory of quadrilateral lattices and
discrete deformations of the algebras (150) are strongly interconnected.
This connection provides us with some new interpretations of notions used in
discrete geometry. For instance, the notion of consistency around the cube
discussed in [70,96] is strickly related to the condition of associativity
\begin{equation}
\mathbf{P}_{1}(\mathbf{P}_{2}\mathbf{P}_{3})\left| \Psi \right\rangle =%
\mathbf{P}_{2}(\mathbf{P}_{1}\mathbf{P}_{3})\left| \Psi \right\rangle =%
\mathbf{P}_{3}(\mathbf{P}_{1}\mathbf{P}_{2})\left| \Psi \right\rangle
\end{equation}
for the three-dimensional algebra of the type (150). Multi-dimensional
consistency ( see e.g. [43]) is the associativity condition (152), i.e.

\begin{equation}
p_{l}(p_{j}p_{k})\left| \Psi \right\rangle =p_{j}(p_{k}p_{l})\left| \Psi
\right\rangle ,\quad j,k,l=1,...,N
\end{equation}
with distinct indices j,k,l. Due to the connection between the
multidimensional consistency and some incidence theorems demonstrated in
[97] the latter are also related to associativity conditions.

\ Discrete Darboux system (217) governs deformations of generic structure
constants for the algebra (150) modulo gauge transformations. Deformations
of the constrained structure constants are of interest too. One of the
examples is provided by the deformations of the three-dimensional algebra
(163)- (165) with the additional constraint

\begin{equation}
L=M=N=1,
\end{equation}
i.e.

\begin{equation}
A+B=0,\quad C+D=0,\quad E+G=0.
\end{equation}
In this case the CS (166)- (169) becomes

\begin{equation}
\frac{A_{3}}{A} =\frac{C_{2}}{C}=\frac{E_{1}}{E},
\end{equation}
\begin{equation}
A_{3}C+E_{1}C-E_{1}A-1 =0.
\end{equation}
Equations (224) again lead to the expressions (191) for A,C,E and equation
(225) takes the form

\begin{equation}
\tau _{1}\tau _{23}-\tau _{2}\tau _{13}+\tau _{3}\tau _{12}-\tau \tau
_{123}=0.
\end{equation}
This equation is the Hirota-Miwa bilinear discrete equation for the KP
hierarchy of B type (BKP hierarchy) [92]. So, the Hirota-Miwa equation (226)
together with the formulae (191) describe discrete deformations of the
algebra (163)- (165) under the constraints (222), (223). In contrast to the
KP case these deformations are not isoassociative.

Equations (177) in the BKP case have the form

\begin{equation}
\Phi _{12}=A(\Phi _{1}-\Phi _{2})+\Phi ,\quad \Phi _{13}=C(\Phi _{1}-\Phi
_{3})+\Phi ,\quad \Phi _{23}=E(\Phi _{2}-\Phi _{3})+\Phi .
\end{equation}
Hence,

\begin{equation}
A=\frac{\Phi _{12}-\Phi }{\Phi _{1}-\Phi _{2}},\quad C=\frac{\Phi _{13}-\Phi
}{\Phi _{1}-\Phi _{3}},\quad E=\frac{\Phi _{23}-\Phi }{\Phi _{2}-\Phi _{3}}.
\end{equation}
Substituting these expressions into (224), one gets

\begin{equation}
\frac{(\Phi _{1}-\Phi _{2})(\Phi _{3}-\Phi _{123})}{(\Phi _{2}-\Phi
_{3})(\Phi _{123}-\Phi _{1})}=\frac{(\Phi _{23}-\Phi _{13})(\Phi _{12}-\Phi )%
}{(\Phi _{13}-\Phi _{12})(\Phi -\Phi _{23})}.
\end{equation}

This is the NSF for the Schwarzian BKP hierarchy [98]. Geometrically this
8-points relation has the meaning of the characteristic equation for two
reciprocal triangles on the plane [49]. Considered as the discrete equation
it defines a lattice on the plane consisting of reciprocal triangles. For
more details including connection with the works of J. C. Maxwell see [49].

Finally, we note that the BKP lattice is gauge equivalent to a particular
Darboux lattice. Indeed, performing the gauge transformation $\Phi =g%
\widetilde{\Phi }$ in equations (227) with $g=\frac{\Phi }{\widehat{\Phi }}$
where $\widehat{\Phi }$ is a solution of (227), one obtains the system of
linear equations for $\widetilde{\Phi }$ with coefficients obeying the
relations (170), (172). The Menelaus lattice is, obviously, the reduction of
the Darboux lattice with L=M=N=0. Thus, the discrete Darboux system (217)
plays the central role in the theory of discrete deformations of the
algebras of the type (150).

\bigskip

\textbf{References}

\bigskip

[1] Zabusky \ N J and Kruskal M D 1965 Interaction of solitons in a
collisionless plasma and recurrence of initial state \textit{Phys. Rev.
Lett. }\textbf{15 }240-243

[2] Gardner C S, Green G M, Kruskal M D and Miura R M 1967 Method for
solving the Korteweg-de Vries equation \textit{Phys. Rev. Lett. }\textbf{19 }%
1095-1097

[3] Gerstenhaber M 1964 On the deformation of rings and algebras \textit{%
Ann. Math. }\textbf{79 }59-103

[4] Gerstenhaber M 1966 On the deformation of rings and algebras II \textit{%
Ann. Math. }\textbf{84 }1-19

[5] Lax P D 1968 Integrals of nonlinear equations of evolution and solitary
waves \textit{Comm.Pure Appl.Math. }\textbf{21}467-490

[6] Zakharov V E and Shabat A B 1974 A scheme for integrating the nonlinear
equations of mathematical physics by the method of the inverse scattering
transform I \textit{Funt.Anal. Appl.} \textbf{8} 226-235

[7] Magri F 1978 Simple model of integrable Hamiltonian equation \textit{%
J.Math.Phys. }\textbf{19 }1156-1162

[8] Miura R M (ed) 1976\textit{\ Backlund transformations, the inverse
scattering method, solitons and their applications ( Lecture Notes in
Physics }vol 515 ) ( Berlin: Sprinder-Verlag)

[9] Matveev V B and Salle M A 1991 \textit{Darboux transformations and
solitons }( Berlin: Springer-Verlag)

[10] Witten E 1990 On the structure of topological phase of two-dimensional
gravity \textit{Nucl.Phys.} \textbf{B 340} 281-332

[11] Dijkgraaf R, Verlinde H and Verlinde E 1991 Topological strings in d<1
, \textit{Nucl.Phys.} \textbf{\ B 352 }59-86

[12] Dubrovin B 1992 Integrable systems in topological field theory \textit{%
Nucl.Phys.} \textbf{B 379} 627-689

[13] Dubrovin B. 1996 \ Geometry of 2D topological field theories, Lecture
Notes in Math. \textbf{1620}, 120-348 (Berlin: Springer-Verlag)

[14] Hertling C and Manin Y I 1999 Weak Frobenius manifolds \textit{%
Int.Math.Res.Notices } \textbf{6} 277-286

[15] Manin Y I 1999 \textit{Frobenius manifolds, quantum cohomology and
moduli spaces} (Providence: AMS)

[16] Hertling C 2002\textit{\ Frobenius manifolds and moduli spaces for
singularities} (Cambridge: Cambridge University Press)

[17] Hertling C and Marcoli M (Ed) 2004 \ \textit{Frobenius manifolds,
quantum cohomology and singularities} \ Aspects of Math. \textbf{E36} \ (
Wiesbaden:\ \ \ Friedr. Vieweg \& Sohn )

[18] Givental A 2003 $A_{n-1}$ singularities and nKdV hierarchies \textit{%
Mosc. Math.J. }\textbf{3 }475-505

[19] Konopelchenko B G and \ Magri F 2007\ Coisotropic deformations of
associative algebras and dispersionless integrable hierarchies \textit{%
Commun.Math.Phys. } \textbf{274} 627-658

[20] Konopelchenko B G and Magri F 2007 Dispersionless integrable equations
as coisotropic deformations: extensions and reductions \textit{Theor. Math.
Phys.} \textbf{151} 803-819

[21] Konopelchenko B G 2009 Quantum deformations of associative algebras and
integrable systems \textit{J.Phys.A: Math. Theor. }\textbf{42 }095201

[22] Konopelchenko B G 2008 Discrete, q-difference deformations of
associative algebras and integrable systems arXiv:0809.1938

[23] Konopelchenko B G 2009 Continuous-discrete integrable equations and
Darboux transformations as deformations of associative algebras \ \textit{%
Theor. Math.Phys. }( to appear) arXiv:0809.3852 2008

[24] Konopelchenko B G 2008 On the deformation theory of structure constants
\ for associative algebras arXiv:0811.4725

[25] Novikov S P, Manakov S V , Pitaevski L P and Zakharov V 1980\textit{\
Theory of solitons.The inverse problem method (}New York\textit{: }Plenum
1984)

[26] Ablowitz M J and Segur H 1981\textit{\ Solitons and inverse scattering
transform }( Philadelphia: SIAM )

[27] Ablowitz M J and Clarkson P A 1991 \textit{Solitons, nonlinear
evolution equations and inverse scattering} (Cambridge: Cambridge University
Press )

[28] Konopelchenko B G 1992 \textit{\ Introduction to multidimensional
integrable equations} (New York and London, Plenum Press)

[29] Toda M 1970 Waves in nonlinear lattices \textit{Prog.Theor. Phys.
Suppl. }\textbf{45 }174-200

[30] Manakov S V 1975 Complete integrability and stochastization of discrete
dynamical systems \textit{Sov.Phys. JETP }\textbf{40 }269-274

[31] Flaschka H 1974 \ The Toda lattice I Existence of integrals \textit{%
Phys. Rev. B }\textbf{9 }1924-1925

[32] Ablowitz M J and Ladik J F 1975 Nonlinear differential-difference
equations \textit{J.Math.Phys. }\textbf{16 }598-603

[33] Hirota R 1977 Nonlinear partial difference equations \textit{%
J.Phys.Soc. Japan }\textbf{43 }I 1429-1433; II 2074-2078; III 2079-2089

[34] Levi D 1981 Nonlinear differential difference equations as Backlund
transformations \textit{J.Phys.A: Math.Gen.}\textbf{\ 14 }1083-1098

[35] Date E, Jimbo M and Miwa T 1982 Method for generating discrete soliton
equations \textit{J.Phys.Soc.Japan }\textbf{51 }I 4116-4124: II 4125-4131

[36] Nijhoff F W, Quispel G R W and Capel H W 1983 Direct linearization of
nonlinear difference-difference equations \textit{Phys.Lett. A }\textbf{97 }%
125-128

[37] Wiersma G, Nijhoff F W and Capel H W 1984 Integrable lattice systems in
two and three dimensions \textit{Lecture Notes in Physics }vol \textbf{239 }%
263-302

[38] Nijhoff F W and Capel H 1995 The discrete Korteweg-de Vries equation
\textit{Acta Appl. Mat. }\textbf{39 }133-158

[39] Kuperschmidt B A 1985 Discrete Lax equations and
differential-difference calculus \textit{Asterisque }\textbf{123}

[40] Suris Yu B 2003 \textit{\ The problem of integrable discretization.
Hamiltonain approach }(Basel: Birkhauser)

[41] Zabrodin A 1997 A survey of Hirota's difference equation \textit{%
Theor.Mat.Phys. }\textbf{113 }179-230

[42] Bobenko A I and Seiler R (Ed) 1999 \textit{Discrete integrable geometry
and physics }(Oxford: Cladendon Press)

[43] Bobenko A I and Suris Yu B 2008 \textit{Discrete differential geometry.
Integrable structures }Graduate Studies in Mathematics vol \textbf{98 }%
(Providence: AMS)

[44] Bobenko A I et al (Ed) 2008 \textit{Discrete differential geometry }(
Basel: Birkhauser)

[45] Dynnikov I A and Novikov S P 2003 Geometry of the triangle equation on
two-manifolds \textit{Moscow Math.J. }\textbf{3 }410-438

[46] Bobenko A I, \ Mercat C and Suris Yu B 2005 Linear and nonlinear
theories of discrete analytic functions. Integrable structure and
isomonodromic Green's function \textit{J.reine angew.Math.}\textbf{\ 583 }%
117-161

[47] Novikov S P 2008 New discretization of complex analysis. The Euclidean
and hyperbolic planes arXiv:0809.2963

[48] Konopelchenko B G and Schief W K 2002 Menelaus' theorem, Clifford
configurations and inversive geometry of the Schwarzian KP hierarchy \textit{%
J.Phys.A: Math.Gen. }\textbf{35 }6125-6144

[49] Konopelchenko B G and Schief W K 2002 Reciprocal figures, graphical
statics and inversive geometry of the Schwarzian BKP hierarchy \textit{%
Stud.Appl.Math. }\textbf{109 }89-124

[50] Schief W K 2003 Lattice geometry of the discrete Darboux, KP, BKP\ and
CKP equations.Menelaus and Carnot theorems \textit{J.Nonl.Math.Phys. }%
\textbf{10 }194-208

[51] Konopelchenko B G and Schief W K 2005 Conformal geometry of the
(discrete) Schwarzian Davey-Stewartson II hierarchy \textit{Glasgow Math.J. }%
\textbf{47A }121-131

[52] Rogers C and Shadwick W F 1982 \textit{Backlund transformations and
their applications }( New York: Academic Press)

[53] Rogers C and Schief W K 2002 \textit{Backlund and Darboux
transformations. Geometry and modern applications in soliton theory }(
Cambridge: Cambridge University Press)

[54] Levi D and Benguria R 1980 Backlund transformations and nonlinear
differential-difference equations \textit{Proc.Nat.Acad.Sci.USA }\textbf{77 }%
5025-5027

[55] Ablowitz M J, Kaup D J, Newell A C and Segur H 1973 Method for solving
the sine-Gordon equation \textit{Phys.Rev.Lett. }\textbf{30 }1262-1265

[56] Zakharov V E, Takhtajan L A and Faddeev L D 1974 A complete description
of solutions of the sine-Gordon equation \textit{DAN SSSR }\textbf{219 }%
1334-1337

[57] Backlund A V 1875 Einiges uber Curven und Flachen transformationen
\textit{Lund Univ.Arsckrift }\textbf{10 }

[58] Seeger A and Kochendorfer A 1951 Theorie der Versetzungen in
eindimensionalen. Atomreichen II \textit{Z.Phys. }\textbf{130 }321-336

[59] Wahlquist H D and Estabrook F B 1973 \ Backlund transformation for
solutons of the Korteweg-de Vries equation \textit{Phys.Rev.Lett. }\textbf{%
31 }1386-1389

[60] Ablowitz M J, Kaup D J, Newell A C and Segur H 1974 The inverse
scattering transform - Fourier analysis for nonlinear problems \textit{Stud.
Appl.Math. }\textbf{53 }249-315

[61] Konopelchenko B G 1982 Elementary Backlund transformations, nonlinear
superposition principle and solutions of integrable equations \textit{%
Phys.Lett.A }\textbf{87 }445-448

[62] Nijhoff F W and Capel H 1990 The \ direct linearization approach to
hierarchies of integrable PDEs in 2+1 dimensions I Lattice equations and the
differential-difference hierarchies \textit{Inverse Problems }\textbf{6 }%
567-590

[63] Dorfmann I Ya and Nijhoff F W 1991 On a 2+1 dimensional version of the
Krichever-Novikov equation \textit{Phys.Lett. A }\textbf{157 }107-112

[64] Bogdanov L V and Konopelchenko B G 1998 Analytic-bilinear approach to
integrable hierarchies II Multicomponent KP and 2D Toda lattice hierarchies
\textit{J.Math.Phys. }\textbf{39 }4701-4728

[65] Bogdanov L V and Konopelchenko B G 1999 Mobius invariant integrable
lattice equations associated with KP and 2DTL hierarchies \textit{%
Phys.Lett.A }\textbf{256 }39-46

[66] Chen H H 1974 General derivation of Backlund transformations from
inverse scattering method \textit{Phys.Rev.Lett. }\textbf{33} 925-928

[67] Wadati M, Sanuki H and Konno K 1975 Relationships among inverse method,
Backlund transformations and infinite number of conservation laws \textit{%
Prog.Theor.Phys. }\textbf{53 }419-436

[68] Gerdjikov V S and Kulish P P 1979 Derivation of Backlund
transformations in the formalism of inverse scattering method \textit{%
Theor.Math.Phys. }\textbf{39 }69-74

[69] Quispel G R, Nijhoff F W, Capel H and van der Linden 1989 Linear
integrable equations and nonlinear difference-difference equations \textit{%
Physics A }\textbf{2-3 }344-380

[70] Nijhoff F W 2002 Lax pair for the Adler ( lattice Krichever-Novikov)
system \textit{Phys. Lett.A }\textbf{297 }49-52

[71] Zakharov V E and Manakov S V 1985 Construction of multidimensional
integrable nonlinear systems and their solutions \textit{Fukt. Anal. Appl. }%
\textbf{19 }11-25

[72] Konopelchenko B G 1993 \textit{Solitons in multidimensions }(
Singapore: World Scientific)

[73] Bogdanov L V 1994 Generic solutions for some integrable lattice
equations \textit{Theor.Math.Phys. }\textbf{99 }505-510

[74] Bogdanov L V and Konopelchenko B G 1995 Lattice and q-difference
Darboux-Zakharov-Manakov system via $\overline{\partial }$-dressing method
\textit{J.Phys.A: Math.Gen. }\textbf{28 }L173-L178

[75] Berezin F A and Perelomov A M 1980 Group theoretical interpretation of
the Korteweg-de Vries type equations \textit{Commun.Math.Phys. }\textbf{74 }%
129-140

[76] Drinfeld V G and Sokolov V V 1981 Korteweg-de Vries type equations and
simple Lie algebras \textit{DAN SSSR }\textbf{258 }11-16

[77] Drinfeld V G and Sokolov V V 1984 Lie algebras and equations of
Korteweg-de Vries type \textit{Current Problems in Mathematics }\textbf{24 }%
81-180

[78] Jimbo M and Miwa T 1983 Solitons and infinite dimensional Lie algebras
\textit{Publ.Res. Inst.Math.Sci.Tokyo }\textbf{19 }943-1001

[79] Marchenko V A 1988 \textit{Nonlinear equations and operator algebras }(
Netherlands: Kluwer)

[80] Mikhailov A V and Sokolov V V 2000 Integrable ODEs on associative
algebras \textit{Commun.Math.Phys. }\textbf{211 }231-251

[81] Van der Waerden B L 1971 \textit{Algebra }( Berlin: Sprinder-Verlag)

[82] Willmore T J 1993 \textit{Riemann geometry }( New York: The Clarendon
Press, Oxford University Press )

[83] Dimakis A and Muller-Hoissen F 1999 Discrete Riemannian geometry
\textit{J.Math.Phys. }\textbf{40 }1518-1548

[84] Novikov S P 2003 Discrete connections on the triangulated manifolds and
difference linear equations arXiv:math-ph 0303035

[85] Wu Ke and Zao Wei-Zhong 2006 Difference discrete connection and
curvature on cubic lattice \textit{Science in China Series A:Math }\textbf{%
49 }1458 ; arXiv: 0707.3741

[86] Zakharov V E 1974 On stochastization of one-dimensional chain of
nonlinear oscillators \textit{Sov.Phys. JETP }\textbf{35 }908-914

[87] Losev A and Manin Yu I 2004 Extended modular operads \textit{Frobenius
Manifolds, Quantum Cohomology and Singularities ( Aspects of Math. }vol 36 )
ed C Hertling and M M marcoli ( Wiesbaden: Vieweg) pp 181-211

[88] Darboux G 1972 \textit{Lecons sur la Theorie generale des surfaces }vol
II ( New York: Chelsea Publ.Comp.)

[89] Doliwa A 1997 Geometric discretization of the Toda system \textit{%
Phys.\ Lett.A }\textbf{234 }187-192

[90] Pedoe D 1970 \textit{Geometry. A comprehensive course} ( Cambridge:
Cambridge University Press)

[91] Brannan D A, Esplen M F and Gray J 1999 \textit{Geometry} ( Cambridge:
Cambridge University Press)

[92] Miwa T 1982 On Hirota's difference equation \textit{Proc.Japan Acad.}%
\textbf{\ 58A }8-11

[93] Witczynski F 1995 On some generalization of the Menelaus theorem
\textit{Zeszyty Nauk Geom.}\textbf{21 }109-111

[94] Doliwa A and Santini P M 1997 Multidimensional quadrilateral lattices
are integrable \textit{Phys.Lett.A }\textbf{233 }365-372

[95] Konopelchenko B G and Schief W K 1998 Three-dimensional integrable
lattice in Euclidean space.Conjugacy and orthogonality \textit{Proc.Royal
Soc. London A }\textbf{454 }3075-3104

[96] Adler V E, Bobenko A I and Suris Yu B 2003 \ Classification of
integrable equations on quad-graphs. The consistency approach \textit{%
Commun.Math.Phys. }\textbf{233 }513-543

[97] Adler V E 2006 Some incidence theorems and integrable discrete
equations \textit{Discrete Comput. Geom. }\textbf{36 }489-498\textbf{\ }

[98] Nimmo J J C and Schief W K 1997 Superposition principles associated
with the Moutard transformation: An integrable discretization of a
2+1-dimensional sine-Gordon equation \textit{Proc.Roy.Soc.London A}\textbf{\
453} 255-279

\end{document}